# Prospects for Detecting Gaps in Globular Cluster Stellar Streams in External Galaxies with the Nancy Grace Roman Space Telescope

Christian Aganze[1,2] , Sarah Pearson[3,9] , Tjitske Starkenburg[4] , Gabriella Contardo[5] , Kathryn V. Johnston[6,7] ,
Kiyan Tavangar[7] , Adrian M. Price-Whelan[6] , and Adam J. Burgasser[2,8]

[1] Kavli Institute for Particle Astrophysics & Cosmology, Stanford University, Stanford, CA 94305, USA
[2] Center for Astrophysics and Space Sciences (CASS), University of California, San Diego, La Jolla, CA 92093, USA
[3] Center for Cosmology and Particle Physics, Department of Physics, New York University, 726 Broadway, New York, NY 10003, USA
[4] Center for Interdisciplinary Exploration and Research in Astrophysics (CIERA), Northwestern University, 1800 Sherman Ave, Evanston, IL 60201, USA
[5] International School for Advanced Studies (SISSA), Via Bonomea, 265, I-34136 Trieste, TS, Italy
[6] Center for Computational Astrophysics, Flatiron Institute, 162 Fifth Avenue, New York, NY 10010, USA
[7] Department of Astronomy, Columbia University, 550 West 120th Street, New York, NY 10027, USA
[8] Department of Astronomy and Astrophysics, University of California, San Diego, La Jolla, CA, 92093, USA
Received 2023 May 19; revised 2023 December 6; accepted 2023 December 12; published 2024 February 15

## Abstract

Stellar streams form through the tidal disruption of satellite galaxies or globular clusters orbiting a host galaxy. Globular cluster streams are exciting since they are thin (dynamically cold) and therefore sensitive to perturbations from low-mass subhalos. Since the subhalo mass function differs depending on the dark matter composition, these gaps can provide unique constraints on dark matter models. However, current samples are limited to the Milky Way. With its large field of view, deep imaging sensitivity, and high angular resolution, the upcoming Nancy Grace Roman Space Telescope (Roman) presents a unique opportunity to increase the number of observed streams and gaps significantly. This paper presents a first exploration of the prospects for detecting gaps in streams in M31 and other nearby galaxies with resolved stars. We simulate the formation of gaps in a Palomar 5–like stream and generate mock observations of these gaps with background stars in M31 and foreground Milky Way stellar fields. We assess Roman's ability to detect gaps out to 10 Mpc through visual inspection and the gap-finding tool `FindTheGap`. We conclude that gaps of $\approx 1.5$ kpc in streams that are created from subhalos of masses $\geqslant 5 \times 10^6\ M_\odot$ are detectable within a 2–3 Mpc volume in exposure times of 1000 s to 1 hr. This volume contains $\approx 150$ galaxies, including eight galaxies with luminosities $> 10^9\ L_\odot$. Large samples of stream gaps in external galaxies will open up a new era of statistical analyses of gap characteristics in stellar streams and help constrain dark matter models.

*Unified Astronomy Thesaurus concepts:* Stellar streams (2166); Dark matter (353); Andromeda galaxy (39); Space telescopes (1547); Galaxy structure (622)

## 1. Introduction

Large-scale cosmological simulations with cold dark matter (i.e., the Lambda cold dark matter (ΛCDM) model) predict hierarchical formation of dark matter halos and the existence of substructure at all scales (White & Rees 1978; Blumenthal et al. 1984; Bullock et al. 2001; Springel et al. 2008; Fiaccom et al. 2016). To test ΛCDM predictions at small scales, previous studies have uncovered satellite galaxies around the Milky Way and dwarf galaxies in the Local Group with stellar masses down to $10^3\ M_\odot$ (Willman et al. 2005; Simon & Geha 2007; Martin et al. 2008; Koposov et al. 2009; Willman et al. 2011; McConnachie 2012; Bechtol et al. 2015; Drlica-Wagner et al. 2015; Geha et al. 2017; Mao et al. 2021). However, in ΛCDM models, galaxies with halos of masses $\lesssim 10^8\ M_\odot$ are faint and more dominated by dark matter compared to higher-mass galaxies, which makes their detection difficult (Efstathiou 1992; Bullock et al. 2000; Okamoto et al. 2008; Sawala et al. 2016). Other dark matter models differ from

ΛCDM in their predictions for the masses and number densities of dark matter subhalos (subhalo mass functions). For instance, warm dark matter (WDM) models (Bode et al. 2001) predict a similar hierarchical collapse at large scales, but this collapse is strongly suppressed at lower masses ($\lesssim 10^9\ M_\odot$, depending on particle mass), resulting in a smaller fraction of low-mass subhalos (Bose et al. 2017). Similarly, some fuzzy CDM models (Hu et al. 2000; Hui et al. 2017), predict a sharp cutoff at low masses ($\lesssim 10^7\ M_\odot$). Self-interacting dark matter (SIDM) models produce halos with pronounced cores with different tidal evolution, masses, and densities compared to CDM halos (Spergel & Steinhardt 2000; Rocha et al. 2013; Tulin & Yu 2018; Forouhar Moreno et al. 2022; Glennon et al. 2022). Even in ΛCDM simulations, the survival and the properties of low-mass subhalos within a larger halo are poorly understood. The tidal field of the central galaxy, preexisting substructure in the halo, and deviations from a smooth spherical halo density profile can all affect the tidal evolution of accreted subhalos (Garrison-Kimmel et al. 2017). On the other hand, analytical calculations, N-body simulations, and high-resolution hydrodynamical simulations show that the central cores of subhalos are likely to survive for long periods, or even indefinitely (van den Bosch & Ogiya 2018; van den Bosch et al. 2018; Errani & Peñarrubia 2020).

All of these differences between different dark matter models can, in principle, be tested by statistical surveys of nearby low-mass subhalos. The critical challenge is detecting these invisible dark subhalos. Strong gravitational lensing offers an opportunity to validate predictions of dark matter models (Dalal & Kochanek 2002; Amara et al. 2006; Nierenberg et al. 2014; Hezaveh et al. 2016; Nierenberg et al. 2017; Gilman et al. 2019). However, this technique probes all subhalos along the line of sight up to the lensed luminous source, complicating the inference of dark matter properties.

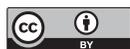

Globular cluster (GC) streams provide a complementary approach for detecting and measuring the spectrum of low-mass subhalos in the Local Volume (Johnston et al. 2002; Yoon et al. 2011; Bovy 2016; Bovy et al. 2017). As GCs orbit the host galaxy, internal evolution and tidal stripping lead to the escape of stars from the central cluster, forming thin, elongated stellar streams which persist for billions of years (Johnston 1998; Helmi & White 1999). A free-floating dark matter subhalo can subsequently perturb these streams, creating a gap-like feature inside the stream (Yoon et al. 2011). Numerical and analytical calculations predict the morphology and frequency of these features in various types of GC streams (Yoon et al. 2011; Carlberg 2012; Erkal et al. 2016; Sanderson et al. 2016; Koppelman & Helmi 2021).

Photometric and spectroscopic surveys have identified and characterized ≈100 stellar streams in the Milky Way, with the majority being GC streams (Odenkirchen et al. 2001; Newberg et al. 2002; Majewski et al. 2003; Newberg et al. 2009; Odenkirchen et al. 2009; Grillmair & Carlin 2016; Mateu et al. 2018; Shipp et al. 2018; Ibata et al. 2019; Li et al. 2022; Martin et al. 2022; Mateu 2023). A few of these GC streams show evidence of gap-like features that are predicted in numerical simulations of dark matter subhalo encounters (de Boer et al. 2018; Bonaca et al. 2020; de Boer et al. 2020; Tavangar et al. 2022). In particular, Price-Whelan & Bonaca (2018) identified a spur and a gap in GD-1, which Bonaca et al. (2019) attributed to a likely encounter with a $10^6$–$10^7$ dark matter subhalo ∼8 Gyr ago, after ruling out other types of perturbers.

Gaps in GC streams can be created through other processes, however. Previous studies have shown that baryonic matter perturbers (galactic bars, molecular clouds, black holes, and spiral arms) can produce similar features in GC streams (Amorisco et al. 2016; Hattori et al. 2016; Price-Whelan et al. 2016; Erkal et al. 2017; Pearson et al. 2017; Banik & Bovy 2019; Bonaca et al. 2020), which makes gaps challenging to decipher, even when they are detected. Moreover, since these streams have intrinsically low surface brightnesses, the detection of gaps has been limited to the Milky Way, which has resulted in relatively small samples.

Detecting gaps in GC streams in external galaxies offers a new window into testing dark matter models by increasing the number and diversity of stream gaps. Previous studies have observed streams in external galaxies arising from tidally disrupted satellites (Martínez-Delgado et al. 2010; Martinez-Delgado et al. 2023). While several candidate GC streams have been proposed in M31 (Pearson et al. 2022), these candidate detections require deeper imaging and spectroscopy to be confirmed, and to map both the streams and gap structures eventually.

The upcoming Nancy Grace Roman Space Telescope (Roman; Spergel et al. 2015) will have a large field of view, high angular resolution, and deep-imaging sensitivity. Pearson et al. (2019, 2022) demonstrated how this combination allows for the detection of very low surface brightness GC streams out to a couple of megaparsecs. In this work, we examine the plausibility of detecting gaps in Palomar 5–like GCs formed from interactions with dark matter subhalos by extending the predictions of Pearson et al. (2019). The paper is arranged as follows: Section 2 describes our methodology for simulating isolated, evolving streams with gaps. Section 3 discusses our simulations of mock observations with Roman including foreground and background star fields, and the feasibility of visual inspection to confirm gaps in simulated star cluster data. Section 4 discusses the application of an automated gap-finding pipeline, FindTheGap (Contardo et al. 2022), to find gaps in simulated data. Section 5 discusses the implications and limitations of this work. We summarize our findings in Section 6.

## 2. Simulating GC Streams with Gaps

Our goal is to simulate observations of gaps in GC streams, starting with the Palomar 5 stream (hereafter Pal 5; Odenkirchen et al. 2001, 2003) as a test case, defined to have a present-day mass of $10^4\,M_\odot$ (Ibata et al. 2017). We conducted all numerical calculations with the gala package (Price-Whelan 2017), which implements numerical integration techniques to model the orbits of stars in a prespecified static potential. We briefly describe our simulation procedure in this section. Additional details on the simulation parameters are given in Appendix A.

### 2.1. Stream Progenitor Coordinates

As no currently known GC streams exist in M31, we used a Pal 5–like stream as an example. In the Milky Way, the Cartesian Galactocentric coordinates of Pal 5 are $(X, Y, Z) =$ (6.1 kpc, 0.2 kpc, 14.7 kpc) and $(V_X, V_Y, V_Z) = (-49.7\ \mathrm{km\,s^{-1}}, -119.4\ \mathrm{km\,s^{-1}}, -11.4\ \mathrm{km\,s^{-1}})$ (Price-Whelan et al. 2019; Vasiliev 2019),[10] and we used identical positions and velocities to simulate a Pal 5–like stream in a M31 potential. Our goal is to test the observability of gaps in streams located at various locations in galactic halos; hence, we also simulated streams at 35 and 55 kpc as the tidal field and the orbit of the stream changes with Galactocentric radius, affecting the stream's physical width. Our choice for the initial coordinates of the stream's progenitor was designed to capture this effect. For simplicity, to simulate equivalent streams at 35 and 55 kpc, we used positions that produce streams at approximately the desired Galactocentric radii, and we assumed that the velocities of the stream at 35 and 55 kpc were the same as the velocity at 15 kpc. We used the same present-day velocities $(V_X, V_Y, V_Z)$ for all streams, but we note that this process results in different orbits compared to the stream at 15 kpc. The initial coordinates and additional parameters of our simulations are summarized in Tables 1 and 2 and in Appendix A.

### 2.2. Generating a Gap in the Stream

We generated model streams using the "particle-spray" method described by Fardal et al. (2015) and implemented in gala. We assumed a uniform mass loss history and a

---

[10] These coordinates are obtained by assuming that our GC progenitor lies at the present-day heliocentric equatorial coordinates of Pal 5, $(\alpha, \delta) = (229°.022, -0°.112)$ at a distance of 20.6 kpc; and has a proper motion vector $(\mu_\alpha \cos\delta, \mu_\delta) = (-2.736\ \mathrm{mas\,yr^{-1}}, -2.646\ \mathrm{mas\,yr^{-1}})$ and radial velocity of $-58.60\ \mathrm{km\,s^{-1}}$. We assumed a Galactocentric coordinate system with the local standard of rest velocity $V_{lsr} = (8.4\ \mathrm{km\,s^{-1}}, 251.8\ \mathrm{km\,s^{-1}}, 8.4\ \mathrm{km\,s^{-1}})$ and the Sun radial distance of 8.275 kpc from the Galactic center based on Schönrich et al. (2010), Bovy et al. (2012), and GRAVITY Collaboration et al. (2019).

**Table 1**
Summary of the Simulation Parameters for the Stream and the Subhalo

|  | Parameter | Description | Range of Values |
|---|---|---|---|
| Stream | $m_p$ | progenitor mass | $5 \times 10^4\ M_\odot$ |
|  | ⋯ | number of particles | ≈80,000 |
| Subhalo | $M_h$ | mass | $2 \times 10^6\ M_\odot$–$10^7\ M_\odot$ |
|  | $r_s$ | scale radius | 0.14–0.32 kpc |
|  | ⋯ | potential | Hernquist[a] |
| Galaxy | ⋯ | potential | Hernquist bulge + Miyamoto–Nagai disk + Navarro–Frenk–White (NFW) halo[b] |

**Notes.**
[a] Hernquist (1990).
[b] Profiles based on Miyamoto & Nagai (1975) and Navarro et al. (1996) with parameters based on Milky Way measurements by McMillan (2017).

progenitor mass ($m_p$) of $5 \times 10^4\ M_\odot$ based on Bonaca et al. (2020). We subsequently simulated the direct impact of a dark matter subhalo on a stream. Throughout these calculations, individual stream stars were treated as noninteracting massless particles, and we did not include the stream's progenitor potential.

To ensure a direct impact between stream and subhalo, we first needed to determine the initial coordinates of both components, given their positions and velocities at the moment of collision. We backward integrated the present-day coordinates of the stream progenitor to time $t_1$, initiated a stream at these coordinates, and then forward integrated by $\Delta t_2$. At this point, the collision position was assumed to be at a position $\Delta x$ away from the progenitor position. We adjusted the coordinates of the subhalo to achieve a fixed relative velocity ($|V_{rel}|$) between the stream stars and the subhalo perpendicular to the impact location. After determining the subhalo's position and velocity at the collision point, we backward integrated its orbit by $\Delta t_2$ again to set the subhalo's initial conditions.

With the initial positions and velocities of the stream and subhalo determined, we forward integrated the system for $\Delta t_2$, computing the stream particle–subhalo interactions using the `gala DirectNBody` routine, with an additional $\Delta t_3$ time to allow the subhalo to pass completely through the stream. At this point, we removed the subhalo from the simulation to avoid potential multiple interactions, allowing for a more direct analysis of the observability of well-defined stream gaps. We then forward integrated the stream stars for the remaining $t_1 - (\Delta t_2 + \Delta t_3)$ to observe the gap's growth over time.

These timescales ($t_1$, $\Delta t_2$, and $\Delta t_3$) were chosen to allow the stream to have similar lengths as that of Pal 5 in M31 (7–12 kpc at $R_{GC} = 15$–55 kpc, based on estimates by Pearson et al. 2019). Additionally, after the subhalo encounter, we continued releasing stars into the mock stream to ensure that there was not a gap at the location of the progenitor. Lastly, we resampled the final stream to match the number of stars observed in Pal 5 in the Milky Way based on Bonaca et al. (2020). Table 2 and Appendix A summarize all parameters for streams at 15 kpc, 35 kpc, and 55 kpc.

### 2.3. Quantifying the Size of the Simulated Gap

To estimate the size of the simulated gap, we fit a Gaussian near the visually identifiable gap. To account for density variations along the stream and the decrease in density near the wings of the stream, we measured both the density ratio and the density difference between the perturbed stream (with a gap) and an equivalent unperturbed stream. By averaging the FWHMs of the Gaussian fits to both the density ratios and density differences, we obtained gap sizes of 1.4 kpc, 1.8 kpc, and 1.8 kpc at $R_{GC} = 15$ kpc, 35 kpc, and 55 kpc, respectively for subhalo masses of $5 \times 10^6\ M_\odot$. We note that in our stream integration procedure, the orbits of the stream and the subhalo, the total integration times, and the impact velocities were chosen to achieve the desired lengths (7–12 kpc) of the stream and to obtain approximately the same gap sizes at all Galactocentric values. Obtaining gaps of comparable sizes at each Galactocentric radius allows us to compare their detection limits systematically. Generally, it is possible to create or smaller larger gaps at these Galactocentric radii by changing the impact parameters of the stream–subhalo encounter (Sanders et al. 2016). We report our parameters in Appendix A.

Figure 1 depicts three simulated Pal 5–like streams at $R_{GC} = 15$ kpc with gaps induced by dark matter subhalos with masses of $2 \times 10^6\ M_\odot$, $5 \times 10^6\ M_\odot$, and $10^7\ M_\odot$. The gap's size increases with the subhalo's mass, as previously shown by analytical and numerical simulations (Yoon et al. 2011; Erkal & Belokurov 2015). Our results are consistent with the numerical simulations of Yoon et al. (2011), who found that gaps induced by $10^5$–$10^{7.5}\ M_\odot$ subhalos can be visually identified in Pal 5–like streams, although they used a higher relative impact velocity ($>100\ km\ s^{-1}$), a single Galactocentric radius ($R_{GC} \approx 25$ kpc), and a longer integration times after the impact (≈4.34 Gyr). Their simulations found that subhalo masses $\geqslant 10^6\ M_\odot$ induce gaps with physical sizes of ≈1 kpc (visually), comparable to the observed gaps in our simulations. However, we note that even when using similar impact parameters and integration times, centrally concentrated halo profiles (e.g., NFW profiles) will result in larger gaps (Sanders et al. 2016).

## 3. Generating Mock Observations of Streams with Gaps in M31 and Other External Galaxies

To model the observability of both streams and gaps, we need to generate mock observations of our streams in external galaxies as they will appear to Roman, by taking into account sensitivity, resolution, and contamination from the Milky Way foreground and the host galaxy halo background stellar populations. To address contaminant populations, we followed a method similar to Pearson et al. (2019) as applied to observations of the halo of M31. However, our approach in simulating gaps and mock observations differs from the method by Pearson et al. (2019) in two main ways: (i) we are now modeling the stream formation process (and the interaction with the dark matter subhalo) and performing this modeling in a more realistic M31 potential; and (ii) we incorporate a model of the Galactic stellar density profile to simulate the Milky Way foreground stars and the stellar halo background stars in M31 and external galaxies. Just like in Pearson et al. (2019), we assumed all stars are resolved down to the magnitude limits.

In Sections 3.1.1 and 3.1.2, we describe our process for generating Milky Way foreground stars and M31 background stellar halo stars; in Section 3.1.3, we estimate the number of stars in a Pal 5–like stream at various galaxy distances; and in

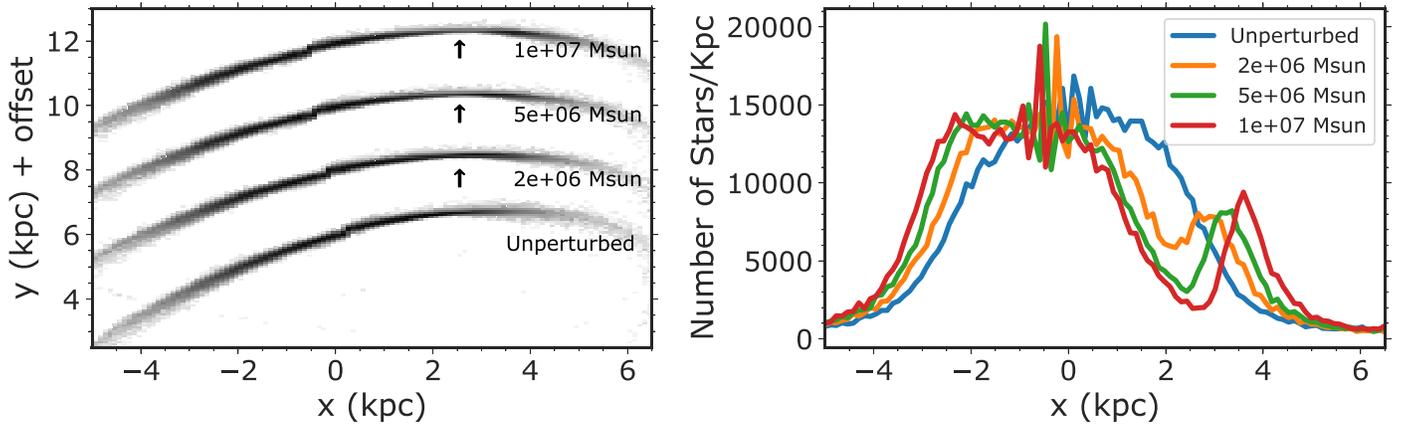

**Figure 1.** Results from our simulations of a gap in a stream at $R_{GC} = 15$ kpc. The total number of stars in each stream is $\approx 80,000$ and the mass of the stream is 50,000 $M_\odot$. Left: Gaps are induced by collisions with dark matter subhalos of masses of 2–$10 \times 10^6 \, M_\odot$. Each stream is offset by a constant displacement in the $y$-direction for display purposes and the relative velocity between the stream stars and the subhalo is 50 km s$^{-1}$. We show an unperturbed stream of the same mass and orbit for comparison. Arrows indicate the location of the gap, and we label the mass of the perturber. Right: Linear density of stars along the $x$-direction in each stream. In both plots, underdensities in perturbed streams can be identified by eye for subhalos with masses $\geqslant 2 \times 10^6 \, M_\odot$.

**Table 2**
Summary of the Stream, Subhalo Coordinates, and Resulting Gap Sizes

| Simulation Step | Parameter and Description | $R_{GC} = 15$ kpc | $R_{GC} = 35$ kpc | $R_{GC} = 55$ kpc |
|---|---|---|---|---|
| initial | progenitor position at $t_1$ ($x_1$, kpc) | (6.1, 0.2, 14.7) | (6.1, 0.2, 34.7) | (6.1, 31.7, 44.7) |
| | progenitor velocity at $t_1$ ($v_1$, km s$^{-1}$) | ($-49.7$, $-119.4$, $-11.4$) | ($-49.7$, $-119.4$, $-11.4$) | ($-49.7$, $-119.4$, $-11.4$) |
| | total integration time ($t_1$, Gyr) | 2 | 3 | 3 |
| collision | $|V_{rel}|$, km s$^{-1}$ | 50 | 70 | 50 |
| | time before collision ($\Delta t_2$, Gyr) | 0.7 | 1.7 | 1.5 |
| | time during collision ($\Delta t_3$, Gyr) | 0.5 | 0.5 | 0.1 |
| | distance of impact from center ($\Delta x$, kpc) | 0.5 | 0.7 | 0.8 |
| result | size of the gap (kpc) | 1.4 | 1.8 | 1.8 |

Section 3.2, we investigate whether these gaps are visible by eye. We summarize our simulation parameters for the Milky Way and M31 stellar populations in Appendix B.

### 3.1. Simulating Mock Observations with Roman

We obtained M31 data from the Pan-Andromeda Archaeological Survey (PAndAS; McConnachie et al. 2009; Martin et al. 2016; McConnachie et al. 2018; Ibata et al., private communication). PAndAS provides wide-field imaging data for the Milky Way, M31, and other nearby galaxies over a total area of 300 deg$^2$, with the 3.6 m Canada–France–Hawaii Telescope (CFHT) MegaPrime/MegaCam camera in the optical and infrared $u$, $g$, $r$, $i$, and $z$ filters. We used extinction-corrected CFHT AB magnitudes (denoted $g_0$ and $i_0$) based on the corrections by Ibata et al. (2014). We selected three patches with projected areas of 10 kpc × 10 kpc at the distance of M31 at radial separations of 15, 35, and 55 kpc from its center, corresponding to regions of $\approx 0.5$ deg$^2$ on the sky. To generate mock Roman observations, we assumed a total field of view of 0.28 deg$^2$ (not simulating the shape of the detector) and limiting Vega magnitudes of $Z$(F087) = 27.15 for 1000 s exposures and $Z$(F087) = 28.69 for 1 hr exposures.[11] As in Pearson et al. (2019), we limited our analysis to the $R$(F062) and $Z$(F087) bands. We did not attempt to simulate any substructures along the line of sight (satellites, streams, or other overdensities; e.g., the PAndAS-MW stream; Martin et al. 2014).

#### 3.1.1. Simulating Milky Way Foregrounds

We simulated foregrounds along the line of sight of M31 assuming a central coordinate of R.A. = 0.57° and decl. = 43°.1 based on the central coordinates of the M31 PAndAS field. We used a Kroupa power-law initial mass function (IMF; Kroupa 2001) for stellar masses between 0.1 $M_\odot$ and 120 $M_\odot$, isochrones from the PAdova and tRieste Stellar Evolution Code (PARSEC; Bressan et al. 2012) spanning ages of 4 Myr to 13 Gyr, and metallicities of $-2.0 \leqslant$ [Fe/H] $\leqslant 0.2$ that realistically encompass the Milky Way thin and thick disks and halo populations. We sampled $10^6$ stars with masses from the IMF, and assigned ages and metallicities based on uniform distributions. We computed the CFHT $g_0$ and $i_0$ and Roman $R$ and $Z$ absolute magnitudes by interpolating in initial mass–absolute magnitude space for every combination of metallicity and age.

We assigned distances drawn from a Galactic density model composed of a thin disk, a thick disk, and a halo based on Jurić et al. (2008; see the parameters in Appendix B). We drew distances from a probability distribution function $P(d) = d^2 \times \rho(R, z)$ out to 100 kpc. After we estimated the distance distribution of Milky Way stars, we computed their observable apparent magnitudes. To model the magnitude uncertainties, we fit the magnitude dependence of the uncertainty ($\delta$mag) for the CFHT $g_0$ and $i_0$ filters based on the McConnachie et al. (2018) point sources.[12] We then assigned apparent $g_0$ and $i_0$ magnitudes for the simulated population by drawing from a normal distribution with a scatter equal to the standard deviation of the estimated dependence. For all Roman magnitudes we assumed a constant uncertainty of 0.1 mag; but the true uncertainties will likely vary with magnitude and exposure time.

Finally, we determined the total number of stars that are observable by Roman at a given magnitude limit by scaling the simulated Milky Way foreground distribution to the observed PAndAS data within the region of the color–magnitude diagram (CMD) bound by $2 < g_0 - i_0 < 3$ and $19.5 < i_0 < 21$. This region in the CMD is predominantly covered by Milky Way foreground isochrones, which makes it ideal for scaling our total number of foreground stars. We then applied the magnitude limit cut corresponding to 1 hr and 1000 s exposures. While this scaling does not consider exact selection effects, it provided a first-order estimate for the number of stars that can be observed by Roman. We obtained agreement between our simulations and both the CMDs and the final $g_0$-band luminosity function from the PAndAS data in Figure 2; we further discuss limitations in our foreground and background simulations in Section 5.2.

### 3.1.2. Simulating Background Stars in M31 and Other External Galaxies

We selected PARSEC isochrone tracks that span ages of 5–13 Gyr and metallicities of $-2.5 \leqslant$ [Fe/H] $\leqslant +0.5$ to cover the approximate range of ages and metallicities of stars in the halo of M31 (Brown et al. 2003; Ibata et al. 2014). Similar to the Milky Way simulation, we assumed a Kroupa IMF for stellar masses between 0.1 $M_\odot$ and 120 $M_\odot$ and a uniform age distribution. We simulated three distinct patches with an area of 10 kpc × 10 kpc at galactocentric radii ($R_{GC}$) of 15, 35, and 55 kpc in the halo of M31. PAndAS and other previous studies (e.g., Escala et al. 2020) have characterized the metallicity and abundance distributions of small regions of M31's stellar halo in detail. To assign metallicities to our simulated M31 stars for each galactocentric radius for a given age, we sampled metallicities from the PAndAS photometric metallicity distribution for M31 stars. We then estimated the absolute magnitude distribution in a given filter using a 2D linear interpolation in log-mass versus metallicity space. Figure 2 compares our simulated and PAndAS metallicities. While the PAndAS metallicities extend below $-2.5$, we were limited by the availability of stellar isochrones below this metallicity. Throughout, we assumed the distance to M31 to be 770 kpc (distance modulus of 24.4; Ibata et al. 2014). To assign distances to stars in the halo of M31, we modeled the stellar density as a flattened spheroid profile based on Ibata et al. (2014; the parameters are discussed in Appendix B). We drew distances assuming that the halo of M31 extends to ≈100 kpc (Chapman et al. 2006) and we assigned projected distances to M31 halo stars as $d = 770$ kpc $+ \bar{z}$, where $\bar{z}$ is the randomly drawn cylindrical galactocentric height for simplicity. Table 3 summarizes other population simulation parameters.

Finally, we assigned apparent magnitudes and magnitude uncertainties similarly as for Milky Way foreground stars. To obtain the correct normalization for the number of stars, we scaled the total number of stars to the observed number between $0.5 < g_0 - i_0 < 2$ and $21.5 < i_0 < 23.5$ in the PAndAS data as there is a significant drop-off in the PAndAS magnitude completeness to below ≈70% for $i_0$, $g_0 > 23.5$ (Martin et al. 2016). As a final check, we examined the simulated luminosity function (number of stars as a function of magnitude) in the CFHT $g$ band based on our CMD-based scaling, luminosity function inferred from PAndAS data.

Figure 2 shows the combined CHFT $i$-band luminosity function of both components (M31 population and the Milky Way foregrounds) compared to the observed luminosity function from the PAndAS data, and it shows the simulated Roman CMD. This figure illustrates that sources with $Z < 25$ dominate Milky Way foreground stellar populations, while M31 includes stars with $Z > 20$. In real Roman observations, it will be possible to separate most Milky Way foreground stars from M31 stars based on their positions on the CMD ($R - Z$ versus $Z$ space, see Figure 2). We note that low-mass stars and brown dwarfs are lacking in our simulated foregrounds; hence they may introduce an additional source of contamination in the real data. Nevertheless, we found general agreement between the simulated and the observed luminosity functions in the CFHT bands. We further compare our simulations and the PAndAS data in Appendix D. While the region corresponding to the Milky Way disk is reasonably well matched, we could not reproduce all of the structures in the CFHT CMDs, perhaps due to an underestimation of the halo and thick disk fraction or our lack of modeling of other density substructures along this line of sight.

To validate further our methodology for simulating the stellar populations of both the Milky Way foregrounds and the M31 stellar halo fields, we compared our surface densities to the CFHT data in the 10 kpc × 10 kpc patches and to the predictions of Pearson et al. (2019). For a 1 hr exposure with Roman, our simulated stellar densities of $8.2 \times 10^4$ stars degrees$^{-2}$ at $R_{GC} = 55$ kpc, $8.2 \times 10^4$ stars degrees$^{-2}$ at $R_{GC} = 35$ kpc and $6.8 \times 10^5$ stars degrees$^{-2}$ at $R_{GC} = 15$ kpc for the halo of M31. These densities are ≈2–6 times higher than the densities obtained by Pearson et al. (2019) at the same radial distances and a similar Roman magnitude cut, but closer to the stellar density profiles fitted to Hubble Space Telescope (HST) results from Brown et al. (2009) to which Pearson et al. (2019) compare. We note that our methods for estimating the foregrounds deviate from the original methodology by Pearson et al. (2019), by incorporating a stellar density model and by scaling the stellar density to the brighter regions of the CMD where PAndAS is most complete. Within the magnitude limits of the PAndAS data ($18 < g_0$ and $i_0 < 26$), our simulated densities of $1.0 \times 10^4$ stars degrees$^{-2}$, $1.0 \times 10^4$ stars degrees$^{-2}$, and $5 \times 10^4$ stars degrees$^{-2}$ at $R_{GC} = 55$ kpc, 35 kpc, and 15 kpc, respectively, are slightly lower (within a factor of 1.4 at 15 kpc and ≈3 at larger $R_{GC}$ values) than the observed densities in PAndAS at the same galactocentric radii. As we scaled the number of stars to a prespecified color range, comparing the stellar density in the PAndAS fields and our simulations provide an additional validation. We further discuss the limitations of our simulations in Section 5.

---

[12] We selected point sources from the McConnachie et al. (2018) catalog by restricting the morphology flags in the $g$ and $i$ bands to $-1$. The catalog can be accessed at https://www.cadc-ccda.hia-iha.nrc-cnrc.gc.ca/en/community/pandas/query.html.

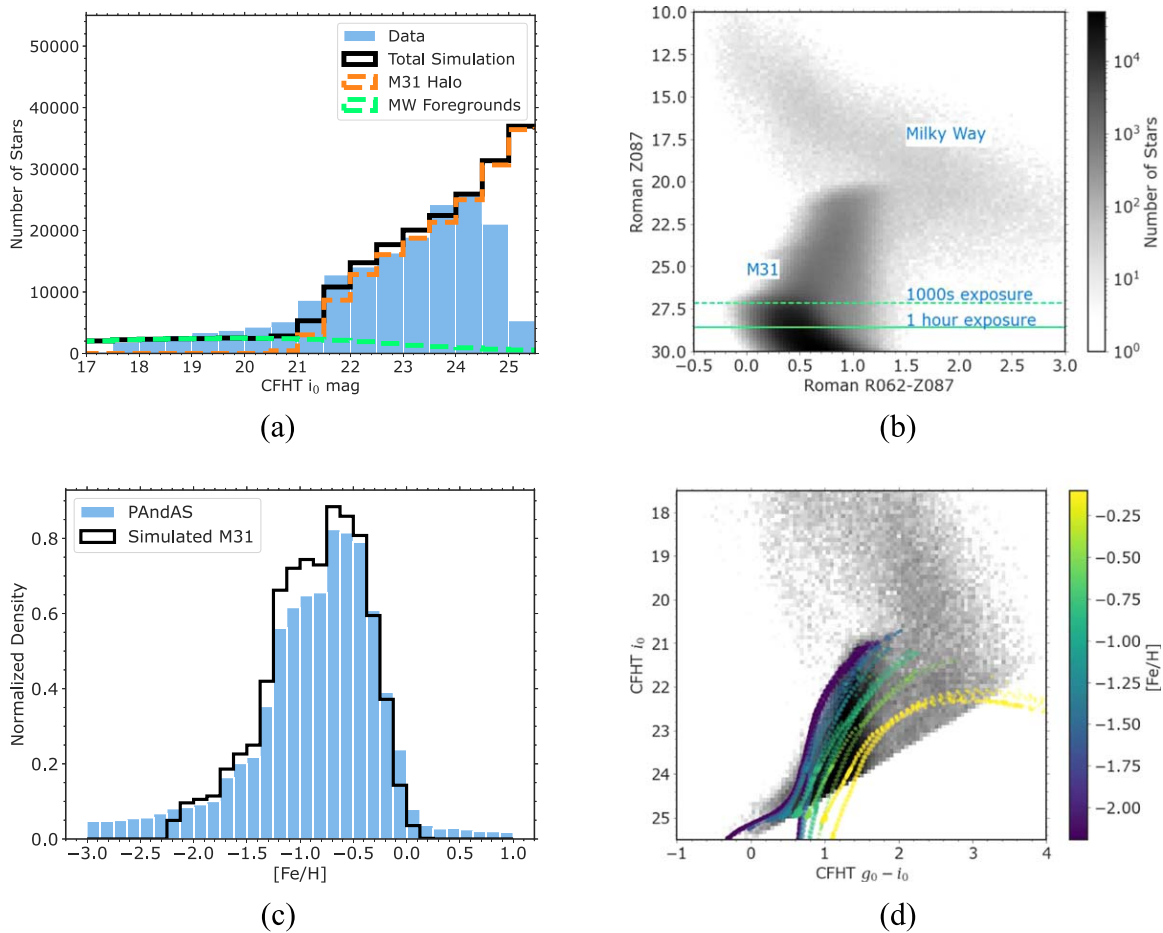

**Figure 2.** Color, magnitude, and metallicity distributions of our simulations compared to the PAndAS data. (a) CFHT $i_0$-mag luminosity function based on PAndAS data (blue filled-in histograms) and our simulations (black) for populations at $R_{GC} = 15$ kpc. The simulations are further divided into Milky Way foregrounds (light green) and M31 stars (orange). We can reproduce the CFHT $i$-band luminosity function based on our scaling to the CFHT CMD (more details are given in Appendix D). (b) Simulated CMD for the Roman $R$ and $Z$ bands at $R_{GC} = 15$ kpc. CMD regions that are dominated by M31 stars or Milky Way foregrounds are labeled in blue text. The horizontal dashed and solid lines show the magnitude cuts for 1000 s and 1 hr exposures, respectively. With these magnitude cutoffs, the M31 halo population will be primarily dominated by horizontal-branch stars and giants. (c) Metallicity distribution of M31 stars in PAndAS in blue at $R_{GC} = 15$ kpc compared to our simulated distribution in black. (d) Simulated CFHT CMD in black, with PARSEC isochrones for stars >5 Gyr color colored by metallicity and offset by the distance modulus of M31.

### 3.1.3. Simulating Observed Stars in Pal 5

We simulated the stream population similarly to the backgrounds, but scaling to the observed properties of Pal 5 in this case. We generated a sample of $10^6$ stars assuming a power-law stellar mass function ($dN/dM \propto M^{-\alpha}$, $\alpha = 0.5$). As argued by the analysis of Ibata et al. (2017), this mass function for Pal 5 can be possibly attributed to the loss of low-mass stars in the cluster before the formation of the stream. We then assigned CFHT $g$ and Roman $R$ and $Z$ absolute magnitudes by interpolating the PARSEC isochrones for an age of 11.5 Gyr and [Fe/H] = −1.3. We applied a distance modulus corresponding to Pal 5 ($d_{mod} = 16.85$; Pearson et al. 2019), and then determined a population normalization factor by comparing the distribution of simulated CFHT $g$ magnitudes to the 3000 stars with $20 \leqslant g \leqslant 23$ that are known members of the Pal 5 stream (Bonaca et al. 2019). With this normalization factor, we computed the number of stream stars detectable in a given host galaxy based on the corresponding distance modulus and Roman magnitude limit. Our number count predictions for Pal 5 match the predictions of Pearson et al. (2019).

To generate a final simulated Roman image, we resampled the simulated streams described in Section 2.2, drawing only the expected number of detectable stars for our Roman $Z$-band limits as a function of distance. We also drew foreground and background stellar fields at the same $Z$-band limits, the latter sampling the three galactocentric radii from M31 PAndAS data and different host galaxy distances. For simplicity, the positions of background and foreground stars were assumed to be uniformly distributed for a given 10 kpc by 10 kpc patch. As most background halo stars are of similar stellar populations as the stellar stream stars, detecting these streams and their gaps depends mostly on density contrasts. We, therefore, focus our analysis on star count maps of these fields, rather than simulated images that incorporate brightness and instrumental point-spread function effects. Additionally, in our analysis, we assume that the stellar stream can be identified using other methodologies, and we focus on the possibility of detecting gaps in the streams. Figure 3 shows examples of our simulations of streams in M31. Henceforth, we will use "mock observations" or "density maps" to describe the results of our simulations.

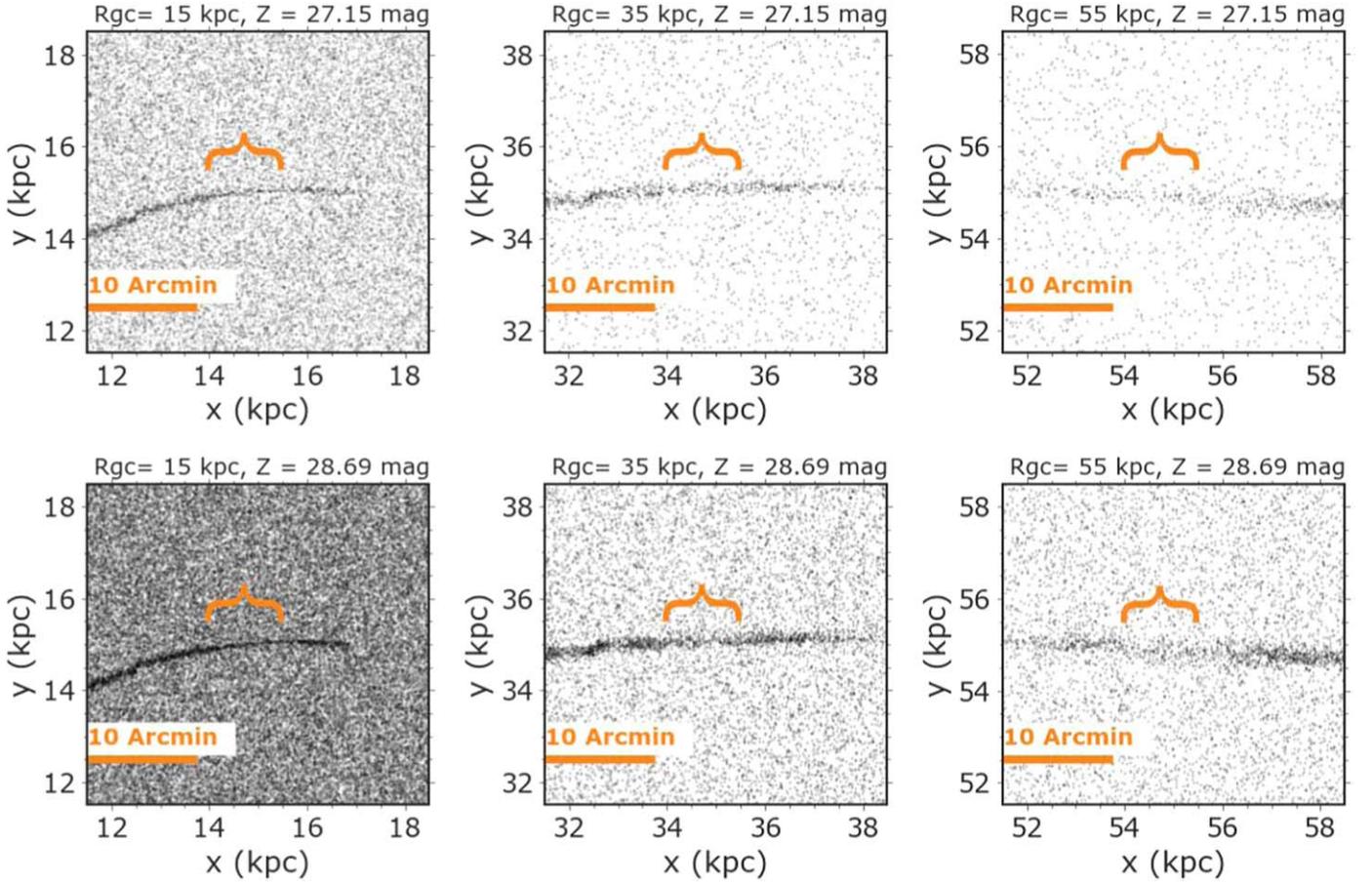

**Figure 3.** Simulated stellar density maps for a full Roman field of view of 0.28 deg$^2$ with M31 background stars and Milky Way foregrounds. The gap in the stellar stream is caused by an interaction with a subhalo with a mass of $5 \times 10^6 \, M_\odot$, and only a portion of the stream is shown here. The streams are injected at galactocentric distances of 15, 35, and 55 kpc, and exposure times are 1000 s ($Z = 27.15$, top panels) and 1 hr ($Z = 28.69$, bottom panels). We display these maps in physical coordinates to highlight the scale of the gap, indicated by curly brackets. The projected $x$- and $y$-coordinates in units of kpc were computed by assuming the distance to M31 is 770 kpc (an angular scale of 13.4 kpc degree$^{-1}$). We can visually see the gap for 1 hr exposure; otherwise, it becomes more obscured by the background population.

**Table 3**
Summary of the Simulation Parameters for Resolved Stellar Populations

| Population | Quantity | Distribution | Reference |
|---|---|---|---|
| Foregrounds & backgrounds | IMF | Kroupa[a] | Kroupa (2001) |
| Milky Way thin disk | age | Uniform (0, 8) Gyr | Jurić et al. (2008) |
| | [Fe/H] | Uniform (−1, 0.5) | Mackereth et al. (2019) |
| | spatial density | Exponential[b] ($H = 350$ pc, $L = 2600$ pc) | Jurić et al. (2008) |
| Milky Way thick disk | age | Uniform (8, 10) Gyr | Kilic et al. (2017) |
| | [Fe/H] | Uniform (−1, 0.5) | Hawkins et al. (2015) |
| | spatial density | Exponential ($H = 900$ pc, $L = 3600$ pc) | Jurić et al. (2008) |
| Milky Way halo | age | Uniform (10, 13) Gyr | Jofré & Weiss (2011) |
| | [Fe/H] | Uniform (−2.5, −1) | Mackereth et al. (2019) |
| | spatial density | Spheroid[c] ($n = 0.64$, $q = 2.77$) | Jurić et al. (2008) |
| M31 halo | age | Uniform (5, 13) Gyr | Ibata et al. (2014) |
| | [Fe/H] | Based on PAndAS data | Ibata et al. (2014) |
| | spatial density | Spheroid[d] ($n = 1.11$, $q = 3$) | Ibata et al. (2014) |
| Pal 5 | age | 11.5 Gyr | Ibata et al. (2017) |
| | [Fe/H] | −1.3 | Ibata et al. (2017) |
| | IMF | $dN/dM = M^{-0.5}$ | Grillmair & Smith (2001) |

**Notes.**
[a] $dN/dM = M^{-\alpha}$, with $\alpha = 1.3$ for masses between 0.08 $M_\odot$ and 0.5 $M_\odot$, $\alpha = 2.3$ for masses > 0.5 $M_\odot$.
[b] Exponential profile defined in Equation B2.
[c] Spheroidal profile defined in Equation B3.
[d] Profile given by Equation B4.

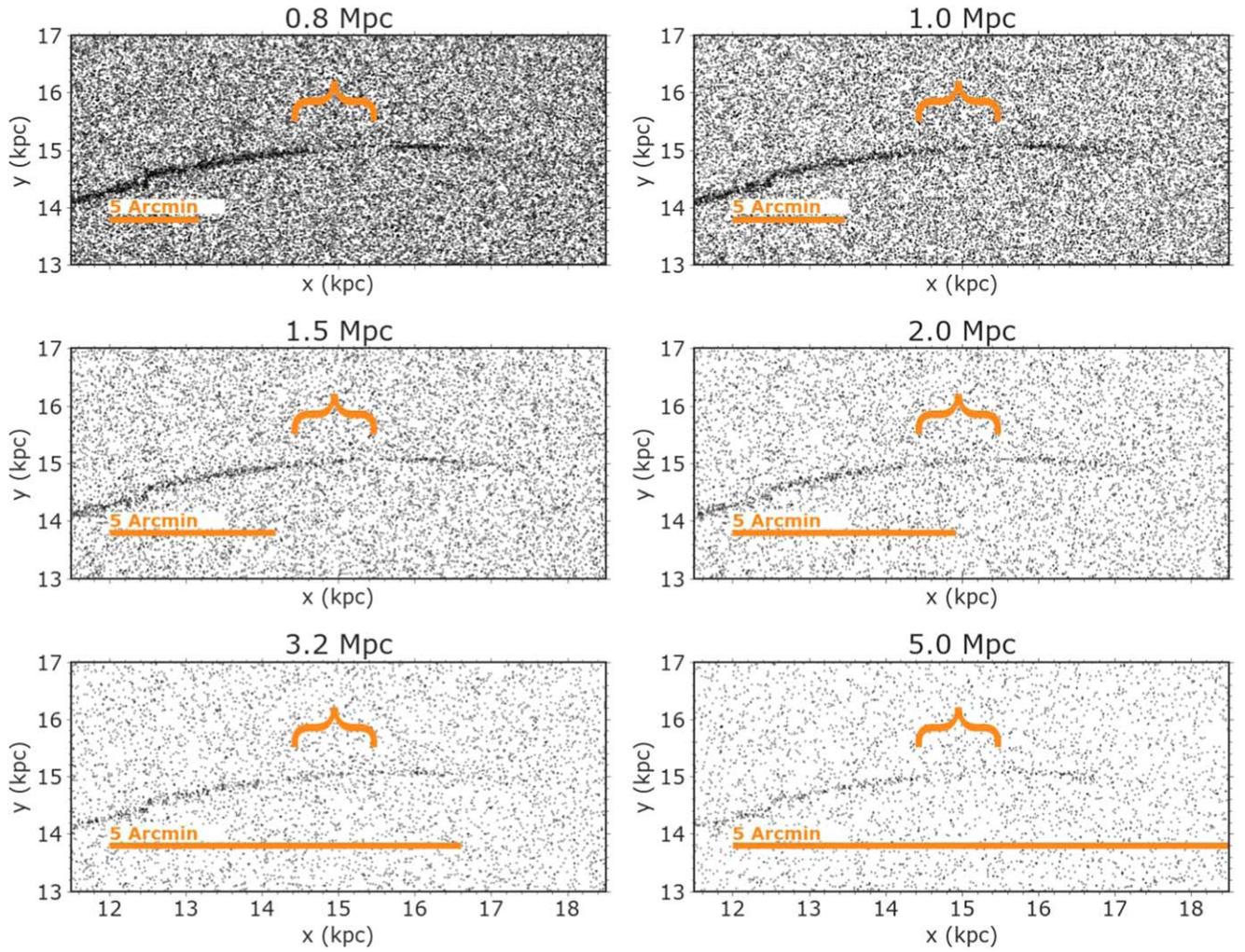

**Figure 4.** Mock observations of streams with gaps in a segment of the full Roman field of view. To generate these mock observations, streams were injected into precomputed backgrounds. Stars are plotted with the same symbol size to facilitate comparison at various galaxy distances. The sizes of the images are 7 kpc × 4 kpc (about half the length of the stream), which would correspond to different angular sizes on the sky depending on the distance of the observed galaxy. The horizontal bar shows the scale of 5′ (or 1/6 of the 32′ × 32′ full Roman field of view). We show the halo of the galaxy at $R_{GC} = 15$ kpc, assuming an exposure time of 1 hr ($Z = 28.69$) and a perturbation in the stream from $5 \times 10^6 \, M_\odot$ subhalo. We can see gaps to a distance of ≈1.5 Mpc.

### 3.2. Gap Identification by Visual Inspection of Density Maps

We now examine gaps and quantify their detection with distance and exposure times. Pearson et al. (2019), Pearson et al. (2022) developed methods for finding Pal 5–like streams in Roman observations. Our primary goal is to investigate the detection of gaps, assuming the stream has already been identified. We present the visual inspection of gaps from simulated streams in M31 (Figure 3) and other external galaxies (Figure 4), centering the density maps on the gap region. For simplicity, we only considered gaps from interactions with subhalos of $5 \times 10^6 \, M_\odot$ as this mass is in the appropriate range for testing different dark matter models (Bullock & Boylan-Kolchin 2017), and gaps that resulted from these interactions are visible in mock streams (Figure 1). In all of our simulated observations, we applied a photometric metallicity constraint of [Fe/H] < −1, as GC streams are typically metal poor (e.g., Martin et al. 2022), allowing us to reduce the number of background stars. In real Roman images, selecting low-metallicity stars will require fitting the foreground populations of a given galaxy to synthetic isochrones.

In Appendix E, we discuss the feasibility of detecting these gaps without requiring this metallicity cut.

The angular lengths of Pal 5 streams for galaxies within a ≈1.2 Mpc volume are larger than the field of view of the Roman telescope; hence, only a portion of the stream will fit inside a Roman field for these distances. For an M31 distance of 770 kpc ($d_{mod} = 24.4$; Ibata et al. 2014), the projected angular distance is 13.4 kpc degree$^{-1}$, making the angular size of Pal 5–like streams in M31 (lengths of 7.8–12 kpc; Pearson et al. 2019) equal or larger than the expected 0°.52 × 0°.52 Roman field of view. Visually, the density contrast between the stream and background stars increases with galactocentric radius and with exposure time. Pearson et al. (2019) estimated that the width of a Pal 5–like stream in M31 would vary between 0.053 and 0.127 kpc at a galactocentric radius ($R_{GC}$) of 15–55 kpc, and the length would vary between 7.8 and 12 kpc. We could best identify the gaps in the density maps for a 1 hr exposure; otherwise, visual identification of the stream and gaps is difficult (see Figure 3).

To simulate streams in other external galaxies with distances spanning 0.5–10 Mpc (assuming a similar stellar composition and tidal field as M31), we offset the M31 background population in the Roman CMD space to the appropriate distance modulus, retaining the same Milky Way foreground populations. The number of stars in Pal 5 was resampled to match the precomputed number of stars at the new galactic distance. We applied the same magnitude cuts as our simulated foreground and background models.

Figure 4 compares the simulated streams, all placed at a fixed galactocentric distance of 15 kpc. We display a fixed area in physical units of 7 kpc × 4 kpc in the figure (7 kpc is ≈half the length of the stream), which corresponds to smaller angular scales in the total Roman field of view at more considerable distances. We show the streams at $R_{GC}$ = 15 kpc because the combination of the density contrast and the thickness of the stream makes it easier to detect the gap in these mock observations visually compared to images at 35 kpc and 55 kpc. As the distance to the host galaxy increases, the density of background stars and the stream decreases. We caution that in actual observations, other external galaxies will have different sizes, and their halos will have a different composition compared to M31. All these caveats are further discussed in Section 5.

Through a visual inspection of streams with gaps in host galaxies spanning a distance of 0.5–10 Mpc, we find that the gap is visible by eye in external galaxies at distances out to ≈1.5 Mpc (see Figure 4).

## 4. Automating the Detection of a Gap

Visual confirmation alone can result in biased assessments of stream and gap detections; hence, we now turn to quantify detection using an automated tool. In Sections 4.1 and 4.2 we lay out methods for defining the gap and the stream region, and in Sections 4.3, 4.4, and 4.5 we outline a procedure for quantifying the detection of each gap. We provide a detection limit as a function of distance using a large sample of simulated mock streams.

### 4.1. Density Estimation and Detecting Gaps

Previous studies have developed algorithms to find and characterize stellar streams in the Milky Way (e.g., Mateu et al. 2017; Malhan & Ibata 2018; Shih et al. 2022, 2023) and in external galaxies (e.g., Hendel et al. 2019; Pearson et al. 2022). Once such algorithms have determined a stream's presence, location, extent, and orientation, we can then search for gaps. We used the gap-finding tool of Contardo et al. (2022), FindTheGap, designed to evaluate underdensities in multi-dimensional data. Gaps, just like streams, can be detected by eye. This tool provides an automated approach and serves as an additional method for confirmation or rejection in conjunction with visual detection.

FindTheGap uses the projection of the second derivatives (Hessian, $\boldsymbol{H}$) of the density estimate onto the orthogonal subspace of the density gradient vector ($g$), denoted as $\Pi\boldsymbol{H}\Pi$, where $\Pi$ is a projection matrix defined as:

$$\Pi = 1 - \frac{gg^T}{g^T g}. \tag{1}$$

The maximum eigenvalue of $\Pi\boldsymbol{H}\Pi$ can then be used as a statistic to estimate if a point in the data space is "in a gap." Conversely, the minimum eigenvalue of $\Pi\boldsymbol{H}\Pi$ can be used to highlight ridges and overdensities. The density estimation depends on a free parameter, the bandwidth, which relates the estimated density to the spacing between data points. In addition to the bandwidth, the stability of the gap detection also depends on the number of data points.

To apply this tool to simulated observations, we started with simulated density maps (Figures 3 and 4), making a cutout centered on the visually identified gap. We did not use the entire Roman field of view as the angular size of the stream becomes progressively smaller at larger galaxy distances, making it more difficult to identify gaps. We then created a grid of 50 by 20 points along each cutout with uniform spacing, covering an area of 5 kpc × 2 kpc that includes the main track of the stream and surrounding foreground and background stars.

The accuracy of this method relies on the choice of bandwidth. Large bandwidths tended to smooth over structures in the data, including the gap, but small bandwidths introduced gaps and other small-scale structures that were not necessarily present in the underlying true density. Additionally, the density estimation in FindTheGap assigns lower densities to regions near the edge of the simulated mock observations. To avoid these edge effects, we first estimated the stellar density and $\Pi\boldsymbol{H}\Pi$ values on a slightly larger data set, incorporating stars beyond the specified grid. Specifically, we required the data bounds to be larger than the grid bounds by a factor of twice the bandwidth. For example, we used a 9 kpc by 6 kpc region for a bandwidth of 1 kpc, given our fixed grid size of 5 kpc by 2 kpc. After we fit the density estimator to the data, we predicted the values of density and $\Pi\boldsymbol{H}\Pi$ on the smaller 5 kpc by 2 kpc grid (see Figure 5). To ensure the fidelity of each gap detection and to remove spurious gaps, we ran this estimation five times for every simulation, choosing the same number of randomly selected stars for each estimation (bootstrap resampling). In each iteration, the density, and the minimum and maximum eigenvalues of $\Pi\boldsymbol{H}\Pi$ were scaled to span values of 0 and 1, respectively, to maintain a consistent range across bootstrap samples. We then computed the final $\Pi\boldsymbol{H}\Pi$ map by taking the median over all bootstraps. Figure 5 shows the result of applying this tool to simulated observations at a distance of 0.8 Mpc. The map of $\Pi\boldsymbol{H}\Pi$ eigenvalues reliably locates the gap and the stream. We further discuss our determination of the optimal bandwidth in Section 4.4.

### 4.2. Further Outlining the Stream and Gap Regions with Indicator Points

In principle, it is possible to determine the stream path from the minimum eigenvalues of the $\Pi\boldsymbol{H}\Pi$ matrix as the stream represents an overdensity. Nevertheless, our focus was solely on detecting gaps within a stream. We constrained the stream region inside the density maps by fitting a second-degree polynomial to the predetermined positions of the injected stream. To define the stream track, we fixed the size of the stream to be 0.2 kpc at $R_{GC}$ = 15 kpc, 0.3 kpc at $R_{GC}$ = 35 kpc, and 0.5 kpc at $R_{GC}$ = 55 kpc. These widths are chosen to encompass the majority of the stars inside the stream visually but exceed the predicted stream width based on the analytical calculations of Pearson et al. (2019) by a factor of 3–5. This process ensures that there are enough grid points to cover the full stream region. This step also allows us to measure the density of stars inside the stream, later discussed in Section 4.5. We note again that we assume that the stream has been observed, and the approximate stream region is therefore known.

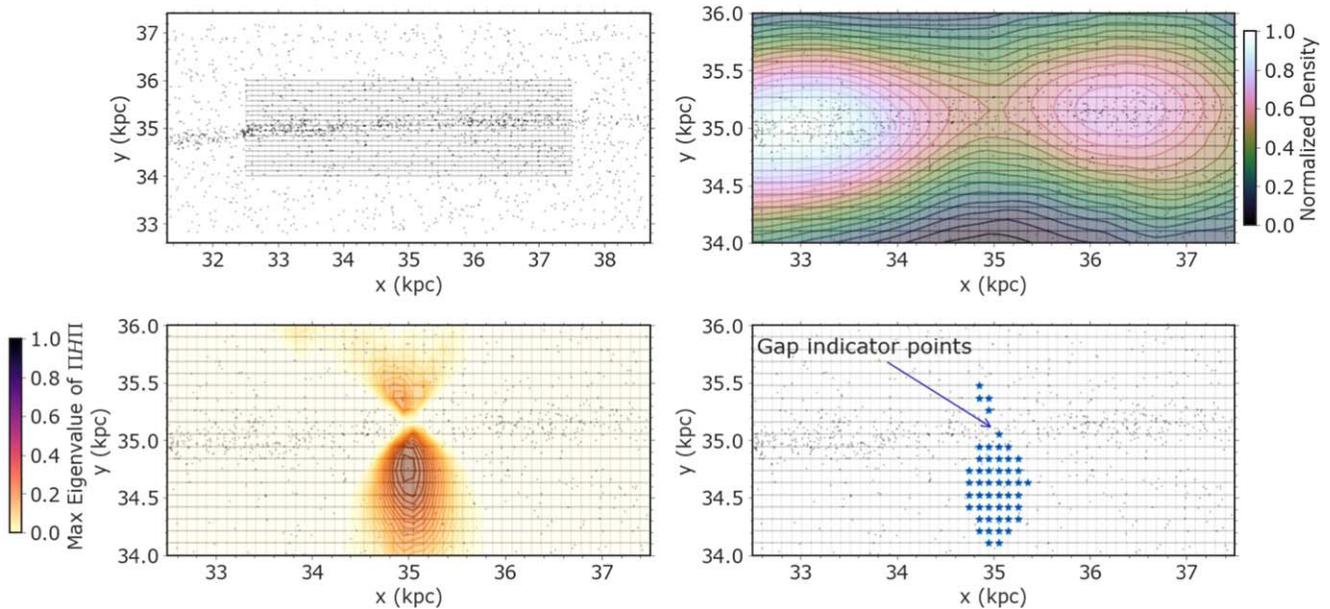

**Figure 5.** Illustration of the gap-detection method with a fixed bandwidth of 0.8 kpc applied to a stream at $R_{GC} = 35$ kpc and a distance of $\approx 0.8$ Mpc for a 1000 s exposure. Note that we assume all stars are resolved. Top left: Black dots show the stream and background stars near the stream. We used a fixed grid of 5 kpc by 2 kpc, as indicated by black lines. Top right: Contours show the stellar density, which clearly shows an underdensity near the gap region and a decrease in density toward the edge of the stream. The density estimation is applied over the full range of the data to avoid edge effects at the end of the grid (see text for discussion). Bottom left: Contours show a map of the maximum eigenvalues of $\Pi H \Pi$, which are maximized near the gap. We used this map to indicate the location of the gap along the stream. Bottom right: Gap indicator points are shown in blue, where the maximum eigenvalues of $\Pi H \Pi$ are in the top 95th percentile. This procedure can locate underdensities inside the stream.

To outline the gap region, we selected points on the grid falling within the top 95th percentile of the distribution of minimum eigenvalues of $\Pi H \Pi$. We will refer to the regions on the grid that match this criterion as "gap indicator points." As shown in Figure 5, this criterion provided a first-order estimation of the location of the gap. We then further constrained the gap region to be centered around the median position of the gap indicator points, with a width equal to the size of the stream region and a length equal to the gap size (see Section 2.3). These definitions of the gap region and indicator points were later used to develop metrics for distinguishing successful detections from noise.

Figure 5 illustrates the gap-detection procedure. As all the mock observations were converted to physical projected coordinates, this assumption yielded consistent gap identification with galaxy distance. We further discuss our quantification of the breakdown of the gap identification procedure in the upcoming sections.

### 4.3. Defining Metrics for Gap Detections

We now discuss applying the gap-detection method on a sample of mock streams in an automated fashion. We created an automatic pipeline using the methods described in Section 4.1 to search for gaps in simulated mock streams, and two metrics to quantify the detections. First, we computed $\Pi H \Pi$ maps and their eigenvalues for mock observations spanning distances between 0.5 and 10 Mpc and using bandwidths between 0.1 and 2 kpc for all three $R_{GC}$ values. We restricted the bandwidth range to 2 kpc as the metrics discussed below did not improve beyond this range. In the end, bandwidths between 0.5 and 0.9 kpc were optimal in finding the gap region. For each step, we repeated the stream generation, the generation of background populations, and the gap-detection process to account for the scatter in the detection metrics at low stellar densities. This process resulted in 538,650 independent mock observations.

We used three metrics to quantify the significance of each detection. To quantify the uncertainty in the gap detection, we defined the spread of all the gap indicator points ($\mathcal{S}_g$) as the range of their $x$-positions (maximum to minimum). As a reminder, "gap indicator points" were defined as points on the grid in the top 95th percentile of maximum eigenvalues of $\Pi H \Pi$. We anticipate that a robust gap detection has low $\mathcal{S}_g$ values, as these gap points would be concentrated around one point near the stream (see the blue markers in the lower right panel of Figure 5).

We then computed the median value of the absolute difference between the $x$-positions of gap indicator points to the center of the density maps, which measures the precision for locating the gap in kiloparsecs, denoted by $\Delta$, computed for each stream separately. As we designed each simulated observation to be centered around the gap, we expect an optimal detection to have a small value for $\Delta$. Nevertheless, in Appendix F.3, we show that our pipeline can also identify gaps away from the center. After we defined $\mathcal{S}_g$ and $\Delta$ per stream, we used these metrics to determine which bandwidths were optimal for detecting gaps.

### 4.4. Determining the Optimal Bandwidth

The effects of bandwidth choice on the gap-detection metrics as a function of distance are shown in Figure 6. We found that bandwidths between 0.5 and 1 kpc resulted in the lowest values for the spread of gap indicator points ($\mathcal{S}_g$) and the deviation of the location of the gap from the center of the density maps ($\Delta$). For large bandwidths, the estimated density on the grid is centrally concentrated and features are washed out. For small bandwidths, the stellar density was fragmented into small groups of spurious gaps, resulting in large values for $\mathcal{S}_g$ and $\Delta$. We illustrate these effects in Appendix F.2. We inferred a middle value of $\approx 0.8$ kpc as our optimal bandwidth.

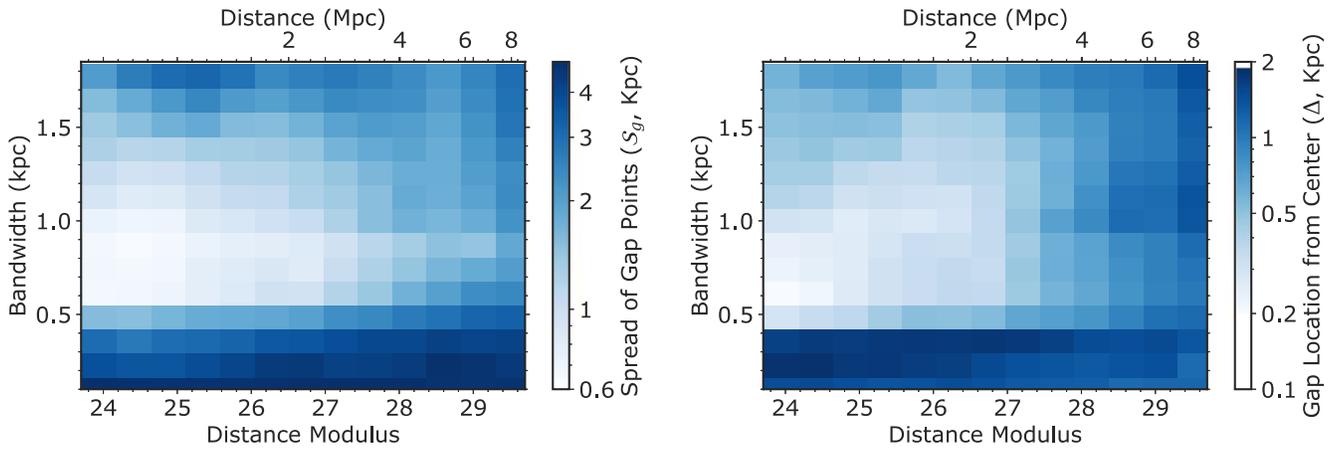

**Figure 6.** Distribution of detection metrics for a gap from a subhalo of mass equal to $5 \times 10^6\,M_\odot$, at $R_{GC} = 35$ kpc and for a 1000 s exposure. Left: Distribution of the spread of gap indicator points ($\mathcal{S}_g$) defined as the range (maximum to minimum) of their $x$-values. There is island of best bandwidths between 0.5 and 1 kpc where this metric is the lowest, which defines our optimal set of bandwidths. Right: Map of the median deviation of gap indicator points with respect to their true locations ($\Delta$). The combination of $\Delta$ and $\mathcal{S}_g$ values show that the optimal bandwidth for locating the gap is between 0.5 and 0.9 kpc.

To evaluate the gap-detection tool's performance further and ensure that our pipeline was robust, we also applied the gap-finding tool to mock observations with an intact stream (without a gap from an interaction with a subhalo) using our described methodology. As shown in Appendix F.3, we could identify the drop-off in density toward the edge of the stream, but we could not identify any gaps inside the track of the unperturbed stream for a bandwidth range of 0.5–0.9 kpc, further validating our method. When applying this tool to real Roman images, the optimal bandwidth may depend on the scale of underdensities in the background. A positive detection would require further characterization. Our goal in this study is to provide an additional methodology for detecting gaps in conjunction with visual inspections. While our pipeline could lead to false positives, it is unlikely to miss any real gaps in the data.

### 4.5. Establishing Tentative Distance Limits for Gap Detections

To determine a tentative detection limit, we used both the density of stars inside the gap region and the location of the gap region as a reference. We estimated the stellar density inside the stream and the gap by counting the number of stars inside each region and dividing this number by the physical area (in units of kpc²; see Section 4.4 for the definition of the gap and stream points/regions). To determine the area of the stream region, we multiplied the total area of the grid (10 kpc²) by the fraction of grid points that fell within each respective region. Because the stream track did not follow a simple straight line, this procedure allowed us to obtain a more accurate measurement of each region's area. For the gap region and the background region, we approximated the area as a rectangle. For the gap, we used a width equal to the width of the stream, and the precomputed length; and for the background, we used a width of 0.5 kpc and a length of 5 kpc.

Figure 7 shows the surface densities (units of number kpc⁻²) inside the stream region, the gap, and the background for bandwidths between 0.5 and 0.9 kpc. As expected, there is a monotonic decrease in the surface density inside the stream, the gap, and the background with increasing galaxy distance. The stream density is generally higher than the gap density and the background. Additionally, streams at smaller $R_{GC}$ values are denser and thinner than streams at larger $R_{GC}$ values. However, it was difficult to determine the detection limits from these densities alone.

To establish a tentative detection limit for our pipeline, we examined the evolution of the gap location with distance. We plot the median value of the location of gap points ($\Delta$) for 10 random streams for each distance step in the bottom panels in Figure 7. We expect the central gap location to remain stable across several iterations for robust detections of gaps. While there was a systematic offset between the center of the stream from the true center, we observed this "flaring" for the value of $\Delta$ at larger distances. While the stellar densities for the background stars are low at $R_{GC} = 55$ kpc, the stream is wide. In contrast, at $R_{GC} = 15$ kpc the stream is thin, which still makes the detection of gaps difficult in both scenarios. At $R_{GC} = 35$ kpc, however, the width of the stream and the stellar background densities are optimal for detecting gaps with our pipeline. By examining the evolution of the gap-detection precision ($\Delta$) with galaxy distance for both 1000 s and 1 hr exposure times, this translates to distance limits of 2–3 Mpc. In Appendix F.1, we discuss examples of gap detections in mock streams at $R_{GC} = 35$ kpc, which show that the location of the gap inside the stream becomes progressively uncertain beyond these distance limits. Additionally, in Appendix E, we demonstrate that gaps are undetectable at $R_{GC} \geqslant 55$ kpc for exposure times $\leqslant 1$ hr with our method if a metallicity cut is not applied to filter out background and foreground stars.

To summarize, we used the tool `FindTheGap` developed by Contardo et al. (2022) to evaluate the detection of gaps beyond a simple visual inspection and to quantify the distance limit with exposure time. We applied this tool to a set of >500,000 mock observations for galaxy distances between 0.5 and 10 Mpc with M31-like stellar populations as background stars. For each mock observation, we defined "gap indicator points" based on gap statistic provided by `FindTheGap`. The gap-detection method relies on the bandwidth as an additional parameter to compute the density of stars on a predefined grid by changing this parameter uniformly between 0.1 and 2 kpc; based on this, we determined that the optimal bandwidth for detecting gaps was between 0.5 and 0.9 kpc. We then evaluated the effectiveness of each detection by estimating the central location of gap indicator points with galaxy distance. The results from this procedure pipeline suggest that gaps from subhalos of $5 \times 10^6 \, M_\odot$ in the halo of M31-like galaxies will be detectable to 2–3 Mpc for exposure times between 1000 s and 1 hr.

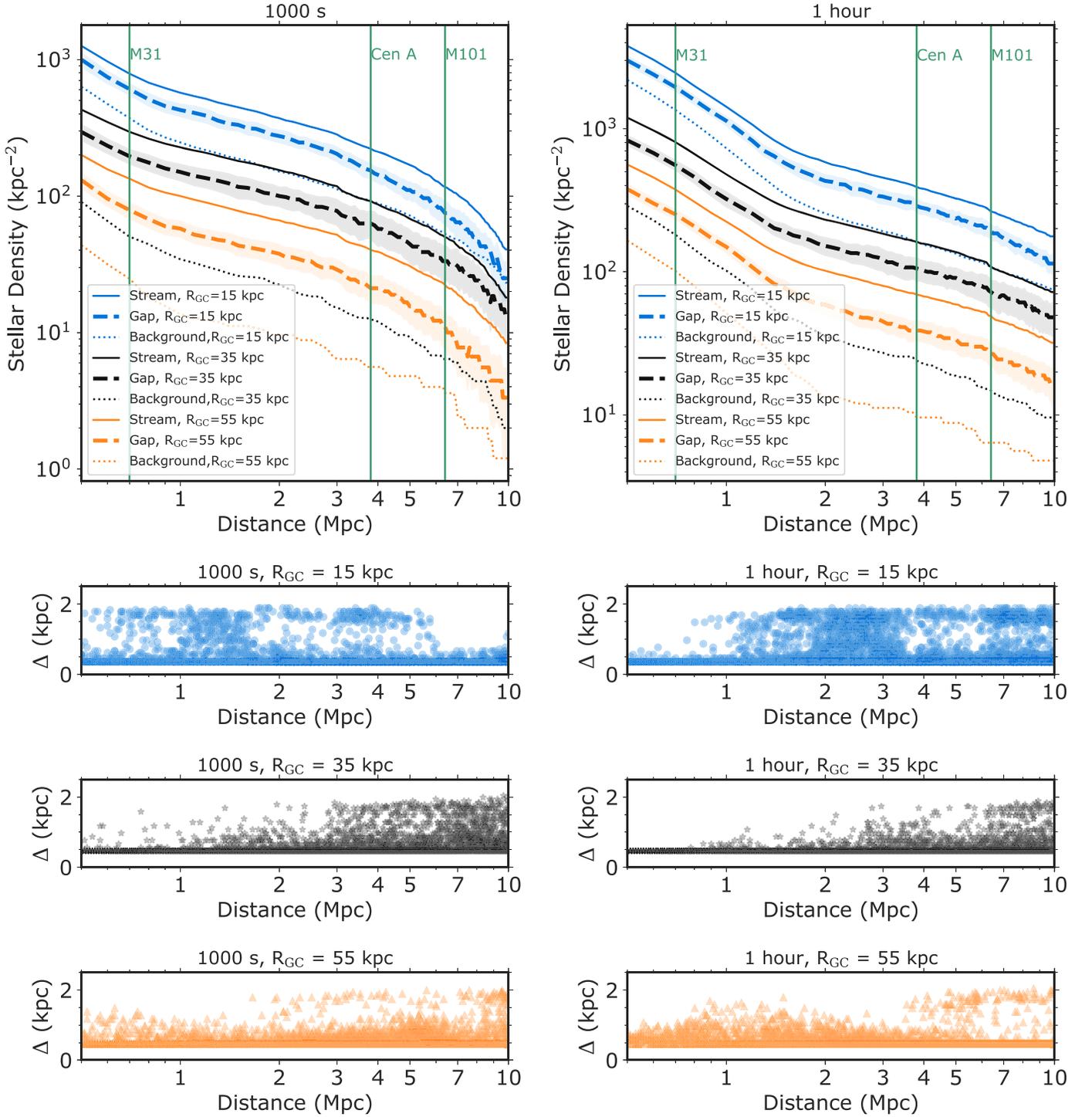

**Figure 7.** Testing the stability of gap detections with distance using the optimal bandwidths of 0.5–0.9 kpc. Gaps were created from subhalos of masses of $5 \times 10^6 \, M_\odot$. First row: Stellar densities in units of kpc$^{-2}$ inside the stream (solid lines), the gap (dashed lines), and the background (dotted lines) with distance for 1000 s (left) and 1 hr exposures (right) for galactocentric radii of 15 kpc (blue), 35 kpc (black), and 55 kpc (orange). Shaded regions show the standard deviation for the measured gap density over 10 samples and vertical lines indicate the distances to nearby bright galaxies. As a general trend, the density inside the stream is higher than the gap and the background density, but it is difficult to establish a detection limit from densities alone. Last three rows: Value of the absolute difference between the $x$-positions of gap indicator points and the center of the density maps ($\Delta$) as a function of distance for a 1000 s exposure (left) and 1 hr exposure (right). The color scheme follows the same pattern as the top panels. We defined our gap-detection limits as where the median value of $\Delta$ fluctuates, corresponding to 2–3 Mpc.

## 5. Discussion

In this section, we revisit assumptions in our numerical simulations (Section 5.1), the observational limitations (Section 5.2), their implications, and how they affect our results. We also discuss prospects of using extragalactic streams for dark matter science (Section 5.3).

### 5.1. Limitations in Our Simulation of a Gap

In our simulations, we have assumed that the galactic potential is smooth and static. However, previous studies of the Milky Way have shown that inhomogeneities in the global potential, including giant molecular clouds, the Galactic bar, streams, other GCs, and spiral arms can perturb GC streams (Amorisco et al. 2016; Hattori et al. 2016; Price-Whelan et al. 2016; Erkal et al. 2017; Pearson et al. 2017; Banik & Bovy 2019; Doke & Hattori 2022). To mitigate this effect, we can search for streams located at large galactocentric radii in external galaxies, where the contrast between the stellar stream and background is more dramatic (see Figure 3). In addition, at large galactocentric radii, bars, spirals, and molecular clouds are less likely to cause dynamical perturbations in streams. Furthermore, with larger sample sizes, we can determine the frequencies, sizes, and locations of underdensities in streams, allowing us to evaluate signatures of perturbations from dark matter subhalos statistically.

Global potentials in galaxies are also deformed by mergers and satellite interactions (Weinberg 1998; Garavito-Camargo et al. 2019), and observations suggest that M31, in particular, has been largely shaped by a possibly recent minor (or major) merger (D'Souza & Bell 2018; Escala et al. 2021; Bhattacharya et al. 2023; Dey et al. 2023). Merger events and interactions with satellites can distort present streams (Erkal et al. 2019; Shipp et al. 2021; Lilleengen et al. 2023) and contribute to the accretion of new GCs that will eventually form streams (Carlberg 2020; Kruijssen et al. 2020; Qian et al. 2022). The details of the formation and disruption of GC streams have not been extensively explored in large cosmological simulations.

Throughout this work, we have only considered one encounter between a stream and a subhalo. Old GC streams can undergo multiple collisions with subhalos, creating multiple observable density fluctuations and perturbations to the stream morphologies (Yoon et al. 2011). In fact, multiple underdensities have been observed in several Milky Way streams such as GD-1 (Bonaca & Hogg 2018) and Pal 5 (Erkal et al. 2017). We do not further explore the effects of multiple encounters here. Still, based on our analysis of the detectability of a gap from a single subhalo encounter, we expect that streams, which have undergone multiple interactions with subhalos to develop multiple observable gaps, can be detected to similar distance limits using our methodology. Using realistic galaxy simulations that include baryonic physics, Barry et al. (2023) predict that Pal 5–like streams in the Milky Way could undergo 2–3 interactions per gigayear with subhalos of masses $>10^6 M_\odot$ before dissolution.

In our analysis, we limited our investigation of the observability of gaps with Roman to subhalo encounters between GC streams and dark matter subhalos with Hernquist profiles (Hernquist 1990). Cuspier profiles for the dark matter subhalo can result in larger gaps for the same encounter properties (Sanders et al. 2016). Additionally, as gaps grow with time, the initial size and the growth of gaps will depend on the collision parameters, such as the mass and scale radius of the subhalo, the relative velocities of the subhalos to the stream, the impact parameter, the stream orbit, the time of the collision, the impact position, and others. These parameters have been extensively explored in numerical and analytical works

(Yoon et al. 2011; Erkal & Belokurov 2015; Sanders et al. 2016; Koppelman & Helmi 2021). These effects will be difficult to disentangle in external galaxies given the lack of kinematic information. However, large statistical sample sizes of observed gaps will allow for rigorous comparisons to predictions of gaps in streams evolved within various dark matter frameworks (e.g., warm, fuzzy, or self-interacting).

### 5.2. Limitations of Our Simulations of Stellar Background and Foregrounds

In this work, we have generated mock Roman observations to mimic stellar halos of external galaxies at various distances without accounting for observational biases due to crowding, extinction, or star–galaxy separation. Pearson et al. (2019) discussed several limitations to consider for such mock observations. In particular, they concluded that crowding effects would not affect the pure detection of thin GC streams in external galaxies with Roman. We can further minimize crowding effects by observing external galaxies with sightlines pointing away from the Milky Way's Galactic plane. Our method relies on estimating the underlying density of stars, as measuring the density contrast between stream stars and background stars is the determining factor in the success of stream detections and of gap detections. Additionally, the effect of dust extinction will be minimal for the halo of M31 at infrared wavelengths (Dalcanton et al. 2015).

Pearson et al. (2019) evaluated the effect of star–galaxy separation on the detection limits of GC streams with Roman. They used the Space Telescope Image Product Simulator to inject known galaxy catalogs into simulated fields and applied quality cuts based on source shape. They concluded that including background galaxies would limit the detection of Pal 5 in M31-like galaxies to 1.1 Mpc (poor star–galaxy separation; no photometric metallicity cut) and 2.6 Mpc ("perfect" star–galaxy separation; applying a metallicity cut of [Fe/H]$- < -1$) for an exposure time of 1 hr. There are $\approx 136$ galaxies in this volume, including seven galaxies with luminosities $> 10^9 L_\odot$ (Karachentsev et al. 2013; Karachentsev & Kaisina 2019), and based on the results presented here gaps will be detectable to these distances. Investigating GC streams 5–10 times more massive than Pal 5, Pearson et al. (2019) estimated that thin GC streams could be detected in host galaxies out to 6.2–7.8 Mpc with a 1 hr Roman exposure for perfect star–galaxy separation and without a metallicity cut (including a metallicity cut they preject the upper limit to 9.3 Mpc). This volume contains $\approx 660$ galaxies where the vast majority within the 7.8 Mpc limit are dwarf galaxies, but 68 galaxies are more luminous than $> 10^9 L_\odot$ and 25 galaxies are more luminous than the Milky Way. While we did not include such an analysis in this paper, because background galaxies will be randomly spread out throughout the image, we expect the feasibility of star–galaxy separation to have a similar effect on the detection of gaps as in the predictions by Pearson et al. (2019) on the detection of streams. We note that the formation of GC streams in dwarf galaxies has been explored in simulations (e.g., Peñarrubia et al. 2009), but more work is needed to estimate their observability.

Our distance limit of $<3$ Mpc contains $\approx 150$ galaxies, including $\approx$eight galaxies with luminosities $> 10^9 L_\odot$ (NGC 55, M32, M31, NGC 300, NGC 404, M33, UGCA 86, and the LMC). Our estimate for the distance limits for the detection of gaps depends on our implicit assumptions that extragalactic halos have similar stellar density fields and metallicity distributions as M31. Additionally, it relies on the presence of a low-metallicity ([Fe/H] < −1) stream population. Dwarf galaxies and the stellar halos of massive galaxies in the Local Group are relatively metal poor (Kirby et al. 2013; Ho et al. 2015), hence a metallicity cut would be appropriate for reducing contamination from Milky Way disk foregrounds, enhancing the stream and gap detection. Additionally, the 3D stellar densities profiles of stellar halos (<50 kpc) for relatively massive (>$10^9$ $L_\odot$) galaxies are consistent with our assumptions ($\rho \propto r^{-n}$, $n$ between −2 and −3; Bullock & Johnston 2005; Johnston et al. 2008; Harmsen et al. 2017; Monachesi et al. 2019; Font et al. 2020). While the results presented in Figure 7 include a M31-like metallicity distribution for the background stellar halo, and a metallicity cut, we have also considered the case of a uniform metallicity distribution for the stellar halo, and present results without a metallicity cut in Appendix E. An M31-like metallicity distribution for stars in the background stellar halo results in a significant fraction of the halo being removed when a metallicity cut is applied. If we instead assume a uniform metallicity distribution for the stellar halo, the modeled M31 stellar halo contains a larger fraction of low-metallicity ([Fe/H] < −1) halo stars, leading to higher background densities after applying the metallicity cut. Nevertheless, using the same requirements for gap detection, we find we can recover gaps out to distances of ∼2 Mpc.

With these considerations, if a Pal 5–like GC stream (i.e., [Fe/H] < −1 with an initial cluster mass ≈ 50,000 $M_\odot$) has formed in the halo of a nearby relatively massive galaxy, the assumptions in our simulations and our applied selection procedure are likely robust.

Finally, future large observing programs dedicated to searching for gaps in streams in external galaxies can extend to longer exposure times, allowing for larger sample sizes and the potential detection of GC streams in dwarf galaxies. Optimistically, it is likely that a full program that is dedicated to observing these gaps with Roman would extend over several hours of observing time, allowing the stacking of images from multiple visits to reach depths beyond our estimates.

### 5.3. Inference of Dark Matter Properties and Expected Sample Sizes

Our focus throughout this paper has been on the detectability of underdensities in extragalactic GC streams. Previous studies have explored pathways to isolate dark matter effects from baryonic effects and infer dark matter properties (e.g., particle mass) from observations of stream gaps. Using linear perturbation techniques, Bovy et al. (2017) constrained the number of dark matter subhalos of masses between $10^{6.5}$ and $10^9$ $M_\odot$ within 20 kpc of the Milky Way's Galactic center by modeling Pal 5 data (see also Banik et al. 2021). Banik & Bovy (2019) provided a powerful method for disentangling underdensities caused by dark matter subhalos from baryonic perturbers (e.g., bars, molecular clouds, and spiral arms) in Pal 5 by computing the various perturbers' contributions to the stream's density power spectrum. They concluded that the contribution from spiral structure to Pal 5's substructure is low, but that giant molecular clouds can create small-scale underdensities comparable to those from dark matter subhalos (see also Amorisco et al. 2016). Recently, Hermans et al. (2021) showed that simulation-based inference techniques with machine learning that map observed densities in streams to simulations can help constrain dark matter structure. They found that GD-1 stream data can constrain WDM particle masses and distinguish between CDM and WDM models. These techniques will be enhanced by the development of fast and accurate stream simulation tools (e.g., Alvey et al. 2023). Lovell et al. (2021) confirmed that the structure in GD-1 and Pal 5 can place limits on the fraction of WDM versus CDM subhalos within a 40 kpc distance from the Galactic center (see also the discussion by Pearson et al. 2019, 2022). Searching for streams in the halos of external galaxies far from the bar and star-forming regions will increase the likelihood of finding gaps induced by gravitational perturbations from dark matter substructure, which can be compared to expectations from various dark matter candidates.

Even though we have shown that Roman will not be able to detect gaps in GC streams in external galaxies further than 2–3 Mpc away for 1 hr exposures, in M31 alone there are ≈450 GCs (Galleti et al. 2006, 2007; Huxor et al. 2008, 2014; Caldwell & Romanowsky 2016; Mackey et al. 2019), which is a factor of three more than the number of known GCs in the Milky Way (Harris 1996, 2010). It is not unreasonable to assume that there is also a factor of three more, yet to be detected, GC streams in M31 than the ≈100 GC streams observed in the Milky Way (Malhan et al. 2018; Martin et al. 2022; Mateu 2023). We know that GCs are also prevalent in other galaxies (Harris et al. 2013). Thus, M31 and other galaxies could provide a diverse set of GC streams with gaps that can be used to constrain substructure within various frameworks of dark matter (Bovy et al. 2017).

While the full survey parameters of Roman are yet to be determined, the proposed High Latitude Survey (HLS) is expected to image high-latitude fields (Akeson et al. 2019). The WFI instrument can reach depths of ≈28 mag (AB) in the $R$, $Z$, and $Y$ bands for exposure times of 1 hr. Furthermore, the instrument has slitless spectroscopic capabilities that cover (0.6–1.8 $\mu$m), which will be beneficial in identifying resolved stellar populations, albeit at a much shallower 1 hr sensitivity. Compared to previous M31 surveys with HST (e.g., the Panchromatic Hubble Andromeda Treasury; Dalcanton et al. 2012), Roman will offer an opportunity to observe M31 at higher efficiency and sensitivity.

Finally, in addition to the Roman, other imaging and astrometric surveys, such as the Vera C. Rubin Observatory, will also help detect new gaps from dark matter subhalos down to ≈$10^6$ $M_\odot$ in dozens of streams in the Milky Way (Drlica-Wagner et al. 2019). These detections will offer the possibility to constrain CDM models at a 99% confidence level, opening up an exciting era for using both Galactic and extragalactic streams to constrain dark matter models.

## 6. Summary

We aimed to quantify the detection prospects of gaps in GC streams in external galaxies with the Roman. To do this, we simulated mock Roman observations of gaps in extragalactic Pal 5–like streams produced by their interaction with dark matter subhalos. We generated mock streams and simulated direct encounters with dark matter subhalos with masses between $2 \times 10^6$ $M_\odot$ and $10^7$ $M_\odot$. Additionally, we simulated realistic mock observations of background of stars in the halo of M31 at galactocentric radii of 15 kpc, 35 kpc, and 55 kpc, taking into account contamination from Milky Way foregrounds. To mimic observations of galaxies at distances that are further than M31, we moved the simulated M31 population to distances of 0.5–10 Mpc, retaining foreground Milky Way populations. To search for gaps in the streams, we first visually inspected mock observations, used the gap-detection tool of Contardo et al. (2022), deriving several metrics to quantify the reliability of our detections with galaxy distance.

We summarize our findings as follows:

1. We find that gaps formed by $5 \times 10^6 \ M_\odot$ subhalos gaps will be visually obvious in 1000 s and 1 hr photometric exposures in the halo of M31.
2. Mock observations of the same stream at various distances from the Milky Way show that gaps can be seen at distances of $\approx 1.5$ Mpc by visual inspection.
3. With the automated detection tool, we confirmed that gaps formed from $5 \times 10^6 \ M_\odot$ subhalos can be identified to distances of 2–3 Mpc, a volume which includes $\approx$ eight galaxies with luminosities $> 10^9 \ L_\odot$.

While our analysis was limited to gaps from in Pal 5–like streams embedded in M31-like halos, it points to the potential of Roman to build a large and diverse set of GC stream gaps in multiple galaxies, which will contribute to constraining various dark matter models.

## Acknowledgments

We thank the CCA dynamics group, Robyn Sanderson, Ivanna Escala, Nora Shipp, and the UCSD Coolstars group for insightful discussions. C.A. thanks Christopher Theissen for providing additional computational resources. This work was developed at the Big Apple Dynamics School in 2021 at the Flatiron Institute. We thank them for their generous support. The Flatiron Institute is supported by the Simons Foundation. This work was made possible through the Preparing for Astrophysics with LSST Program, supported by the Heising-Simons Foundation and managed by Las Cumbres Observatory. Support for S.P. was provided by NASA through the NASA Hubble Fellowship grant #HST-HF2-51466.001-A awarded by the Space Telescope Science Institute, which is operated by the Association of Universities for Research in Astronomy, Incorporated, under NASA contract NAS5-26555. T.S. was supported by a CIERA Postdoctoral Fellowship.

*Software:* Astropy (Price-Whelan et al. 2018), Scipy (Virtanen & Gommers et al. 2020), Matplotlib (Hunter 2007), Seaborn (Waskom et al. 2014), Numpy (Harris et al. 2020), Pandas (McKinney 2010), Scikit-learn (Pedregosa et al. 2011), and FindTheGap (Contardo et al. 2022).

## Appendix A
## Gravitational Potentials and Subhalo Collision Parameters

In our analysis, we used M31 as a model for the external host galaxy. Many groups have estimated the total mass and the potential of M31 by modeling the kinematics of satellites (Watkins et al. 2010); constraining the rotation curve (Chemin et al. 2009); modeling the velocity distributions using tracer particles such as stars, GCs, and planetary nebulae (Kafle et al. 2018); dynamical modeling of the giant southern stream in M31 (Fardal et al. 2013); and the Local Group timing argument (González et al. 2014; Chamberlain et al. 2023). Additional references and trade-offs from these techniques are summarized by Fardal et al. (2013) and Kafle et al. (2018). To set up an M31-like potential, we used the model of Kafle et al. (2018), which is composed of a central bulge, a disk, and a halo. The bulge potential follows a Hernquist profile (Hernquist 1990) given by:

$$\Phi_b(r) = -\frac{GM_b}{r+q},$$  (A1)

with a scale length ($q$) of 0.7 kpc and a bulge mass($M_b$) of $3.4 \times 10^{10} \ M_\odot$. The disk potential follows a Miyamoto–Nagai density profile (Miyamoto & Nagai 1975) given by:

$$\Phi_d(R, z) = -\frac{GM_d}{\left(R^2 + \left(a + (z^2 + b^2)^{\frac{1}{2}}\right)^2\right)^{1/2}},$$  (A2)

with a scale length ($a$) of 6.5 kpc, a scale height ($b$) of 0.26 kpc, and a total disk mass ($M_d$) of $6.9 \times 10^{10} \ M_\odot$. These parameters for the disk and the bulge were adopted from a compilation of literature values (Bekki et al. 2001; Font et al. 2006; Geehan et al. 2006; Seigar et al. 2008; Chemin et al. 2009; Corbelli et al. 2010; Tamm et al. 2012). We assumed an NFW profile (Navarro et al. 1996) for the halo given by:

$$\Phi_h(r) = -\frac{G \ M_{\rm vir} \ln(1 + r \ c/r_{\rm vir})}{g(c) \ r},$$  (A3)

with $g(c) = \ln(1 + c) - c/(1 + c),$  (A4)

$$M_{\rm vir} = \frac{4\pi}{3} r_{\rm vir}^3 \Delta \rho_c,$$  (A5)

and $\rho_c = \frac{3H_0^2}{8\pi G},$  (A6)

where $M_{\rm vir}$ is the virial mass, $r_{\rm vir}$ is the virial radius, $c$ is the concentration parameter, $\Delta$ is the virial overdensity parameter, and $\rho_c$ is the critical density of the Universe. As many of these parameters are interrelated, we used best-fit values for the halo virial mass at $\Delta = 200$ of $M_{200} = 0.7 \times 10^{12} \ M_\odot$ and $\log c = 1.5$ based on the inferred posterior distribution of Kafle et al. (2018). We note that the concentration parameter was poorly constrained in this work. Additionally, while their inferred mass is lower compared to recent literature values ($1–2 \times 10^{12} \ M_\odot$, as summarized by Bhattacharya et al. 2023), we keep this assumption for self-consistency. A recent measurement by Dey et al. (2023) reports a mass of $0.6 \times 10^{12} \ M_\odot$, consistent with our assumption. We also assumed $H_0 = 67.7$ km s$^{-1}$ Mpc$^{-1}$ based on Planck results (Planck Collaboration et al. 2020).

For the dark matter subhalo, we again assumed a Hernquist density profile with masses ($M_h$) and radii ($r_h$) determined by the scaling relation from Erkal et al. (2016):

$$r_h = 1005 \ {\rm pc} \times \left(\frac{M_h}{10^8 M_\odot}\right)^{0.5}.$$  (A7)

Tables 1 and 2 summarize our simulation parameters for the creation of a stream with a gap inside a galaxy potential.

## Appendix B
## Milky Way and M31 Population Simulation Parameters

We modeled the Milky Way density as a three-component model based on Jurić et al. (2008) given by:

$$\rho = \rho_{\text{thin disk}} + f_0 \times \rho_{\text{thick disk}} + f_1 \times \rho_{\text{halo}}, \tag{B1}$$

where $f_0$ and $f_1$ are the relative fraction of thick disk and halo stars to the thin disk population at the position of the Sun, set to 0.12 and 0.005, respectively. Stellar densities for the disk were assumed to follow exponential profiles parameterized by a scale height ($H$) and scale length ($L$):

$$\rho_{\text{disk}} = \rho_\odot \exp\left(-\frac{R - R_\odot}{L}\right) \exp\left(-\frac{|z - Z_\odot|}{H}\right). \tag{B2}$$

For the thin disk, we assumed $H = 300$ pc and $L = 2600$ pc. For the thick disk, we assumed $H = 900$ pc and $L = 3600$ pc. We also assumed $R_\odot = 8.3$ kpc and $Z_\odot = 0.027$ kpc (Jurić et al. 2008). While the scale height of a population varies with its main-sequence lifetime (Bovy 2017), and dynamical evolution leads to asymmetries in the density profile (Reylé et al. 2009; Liu et al. 2017; Nitschai et al. 2021), these simple assumptions provide a reasonable first-order estimate of the broad stellar densities of present-day Milky Way stellar populations. For the halo stellar density, we used a flattened spheroid profile:

$$\rho_{\text{halo}} = \left(\frac{R_\odot}{(R^2 + (z/q)^2)^{\frac{1}{2}}}\right)^n, \tag{B3}$$

with $q = 0.64$ and $n = 2.77$.

The 3D stellar density for the halo of M31 is given by:

$$\rho_{\text{M31}} = \left((\tilde{R}^2 + (\tilde{z}/q)^2)^{\frac{1}{2}}\right)^n, \tag{B4}$$

(see Equation (B3)), where $\tilde{R}$ and $\tilde{z}$ are the cylindrical radius and height starting from the center of M31, in the plane and perpendicular to its disk, respectively, with $q = 1.11$, and $n = -3$.

## Appendix C
## Simulated Streams at 35 and 55 kpc

In Figure 8, we show the simulated streams at larger galactocentric radii, with gaps from perturbations due to subhalos of masses between 2 and $10 \times 10^6\ M_\odot$. The stream and gap sizes for all our simulations are provided in Table 2. Streams at larger galactocentric radii are thicker than the stream at 15 kpc, which affects the feasibility of detecting gaps (see Section 4.5).

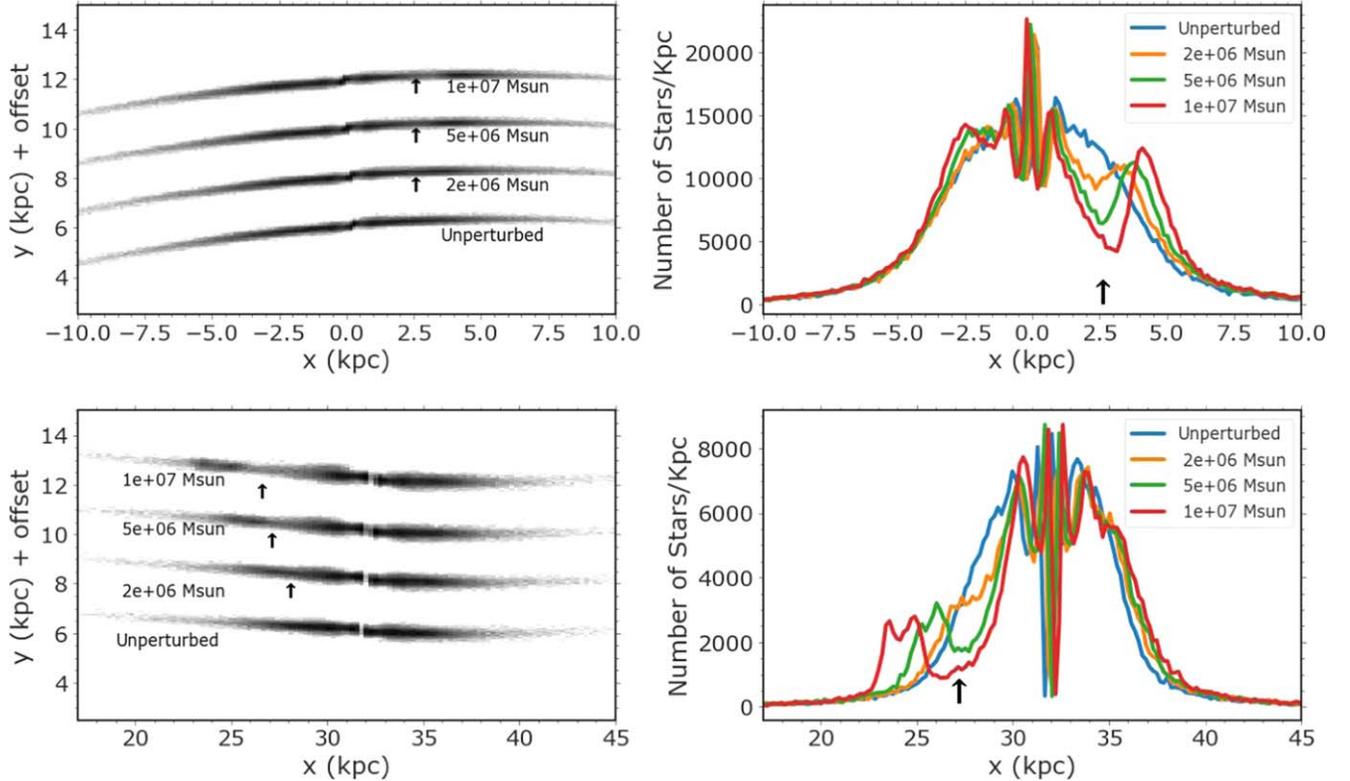

**Figure 8.** Similar to Figure 1. The left panels show the streams at $R_{\text{GC}} = 35$ kpc (top) and $R_{\text{GC}} = 55$ kpc (bottom). Arrows indicate the location of the gap created by subhalos with masses between 2 and $10 \times 10^7\ M_\odot$, with bigger subhalos creating the largest gaps. The right panels show the 1D density of stars along the stream to highlight the gap morphology further.

## Appendix D
## Comparing Simulated Populations to PAndAS data

In Figure 9, we show simulated CMDs in the CFHT $g_0$ and $i_0$ bands compared to the reddening-corrected PAndAS data, which reproduces a significant portion of the range of colors and magnitudes covered by the data. We also show the regions of the CMDs that were used to scale the simulation to the data. While our simulations are a reasonable match to the data, we did not reproduce the overdensities at $g_0 - i_0 \approx 1$ and $i_0 > 22$, which were labeled as Milky Way halo stars by

Ibata et al. (2014), pointing to perhaps an underestimation of the fraction of Milky Way disk to halo stars in our simulations. Additionally, our simulations assume magnitude completeness down to the magnitude limits, which is not valid for real data. As reported by Martin et al. (2016), the completeness of PAndAS drops below 70% for $i_0 > 23$. Nevertheless, as discussed in the main text, this scaling provided a robust estimation of the CFHT $g$-band luminosity function and the total stellar density within the PAndAS fields.

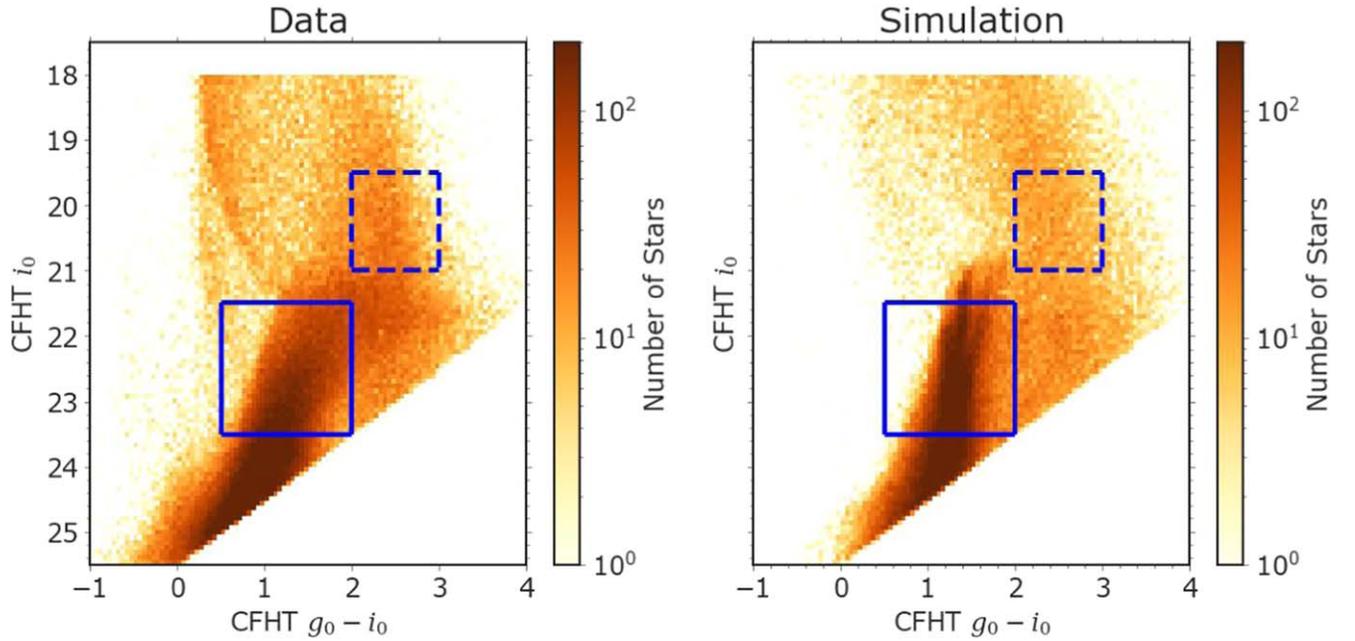

**Figure 9.** Comparison between the simulated stellar populations and the PAndAS data. Left: CMD of the CFHT data from PAndAS covering distances of 10–20 kpc from the center of M31. We used the regions shown by the dashed rectangles to scale the number of the Milky Way foreground stars and the regions shown in solid rectangles to scale the total number of stars in our simulations. Right: Similar to the left panel, but we show the simulated populations here. Our simulated CMD reasonably matches the PAndAS observations.

# Appendix E
# Detection Limits Without a Metallicity Cut

We also report detection limits without applying an additional metallicity cut. To generate mock observations, we followed the same procedure as in our simulations with a metallicity cut, and we applied the `FindTheGap` tool to these mock observations with the optimal bandwidths of 0.5 to 0.9 kpc. Figure 10 shows the resulting stellar densities and our gap-detection metric. Gaps at 35 kpc show the clearest distinction between detections (i.e., where the density inside the gap is above the background and the location of the gap is stable) and nondetections. By visually inspecting the simulated stellar density maps in Figure 10, it is unlikely that the gaps at galactocentric radii of 55 kpc are detectable using our method, but gaps at smaller galactocentric radii will be detected to ≈1 Mpc. While the stellar backgrounds are less dense at larger galactocentric radii, the streams are wider, which makes it hard to detect if the background levels are high.

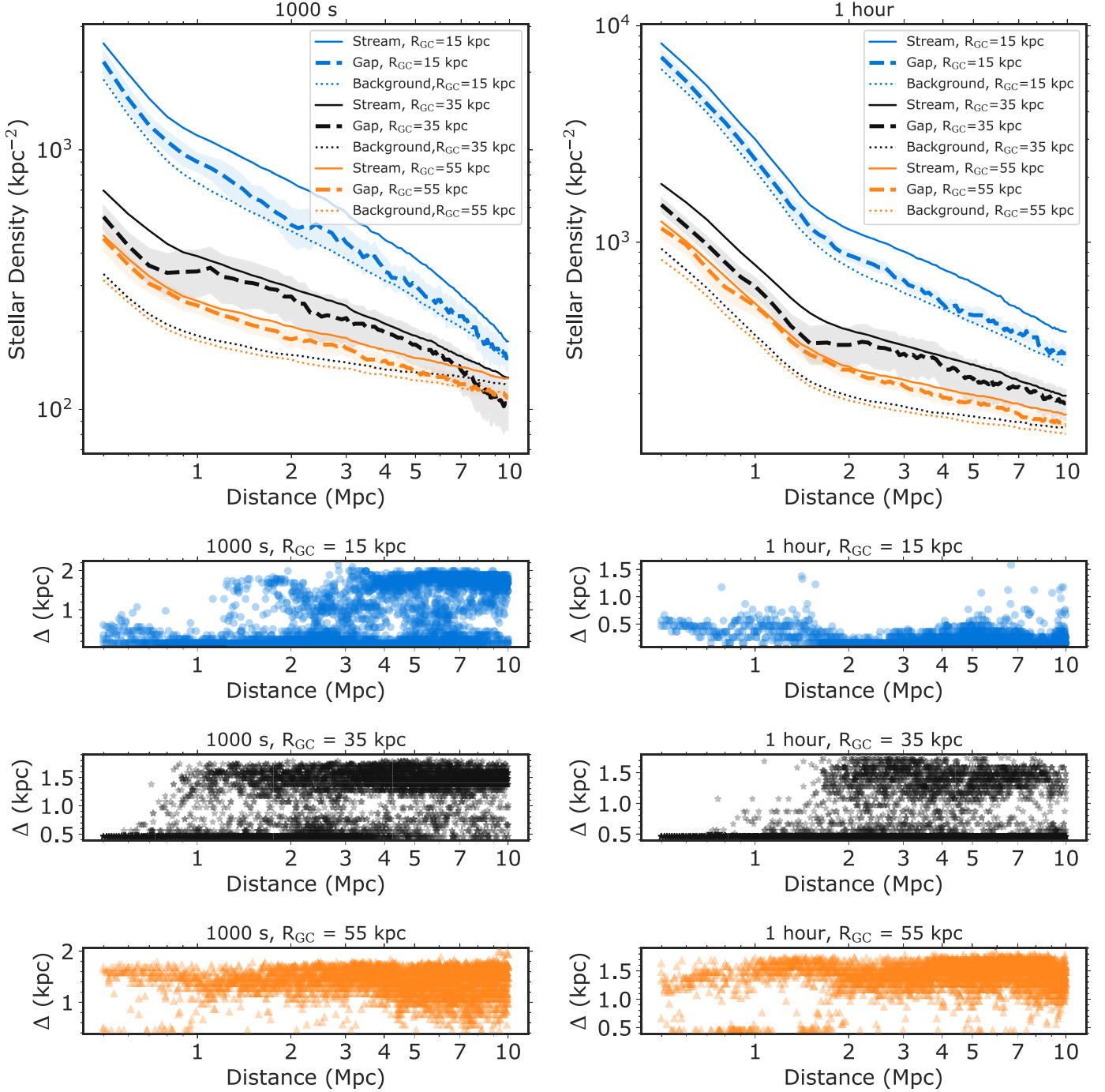

**Figure 10.** Similar to Figure 7 without applying a photometric metallicity cut to the data. The top plots show the median stellar density inside the stream region, the gap region, and the background with shaded regions showing one standard deviation contours. The bottom plots show the location of the gap metric ($\Delta$, see Section 4.3) at different galactocentric radii. Gaps are likely detectable within <2 Mpc for $R_{GC}$ = 35 kpc and 1 hr, but gaps at $R_{GC}$ = 55 kpc are probably undetectable with our method.

## Appendix F
## Additional Checks of the Gap Detection Pipeline

### F.1. Visual Inspections of Gap Detections with Distance

Figure 11 shows additional examples of the density of stars in mock observations as a function of distance and the identification of gaps. Here, we only show streams simulated for a 1000 s exposure time. The localizations of the gaps are shown by arrows, which are stable <2 Mpc for $R_{GC} = 35$ kpc and 55 kpc and <5 Mpc for $R_{GC} = 15$ kpc, but become scattered at larger distances.

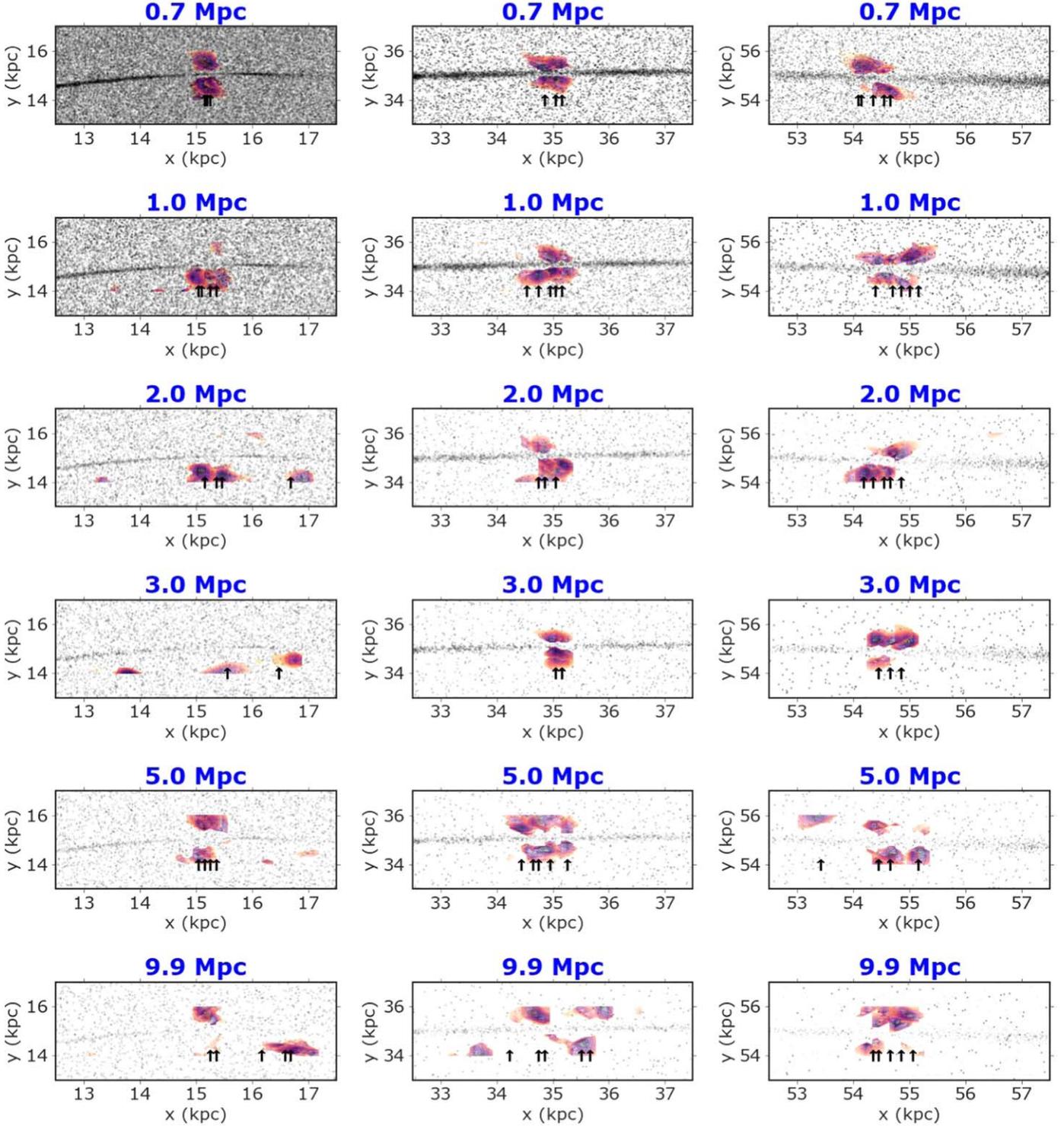

**Figure 11.** Additional simulation of a gap from a $5 \times 10^6 \, M_\odot$ subhalo with increasing distance to the host galaxy (top to bottom) and increasing galactocentric radius (left to right). Each panel is a composite of five mock observations, and all panels are centered around the gap. Contours show the distribution of the top 95th percentile of the maximum eigenvalues of $\Pi H \Pi$ values based on our density estimator with a bandwidth of 0.6–0.8 kpc. Simulated stars are shown as black points. Vertical arrows show the center gap area based on our pipeline for each iteration in a centrally located gap characterized by successful identifications of gaps. Our method successfully identified gaps when the density of stars inside the stream was relatively high (distance < 2 Mpc) for this exposure time. We discuss our characterization of potential failure modes of our pipeline in Appendix F.2.

We demonstrate where the detection of gap breaks down as a function for mock observations $R_{GC} = 15$ kpc by showing three cases in Figure 12: (a) the case where the gap was detectable by eye, but the bandwidth was much smaller than our optimal bandwidth; (b) the case where the gap was detectable by eye and the bandwidth was much larger than our optimal bandwidths; and (c) the case where the bandwidth was optimal, but the stellar density in the stream was extremely low. For the first and second cases, the spread in the location of gap points ($\mathcal{S}_g$) was large. In the last case, the deviation of gap points ($\Delta$) from the center was large.

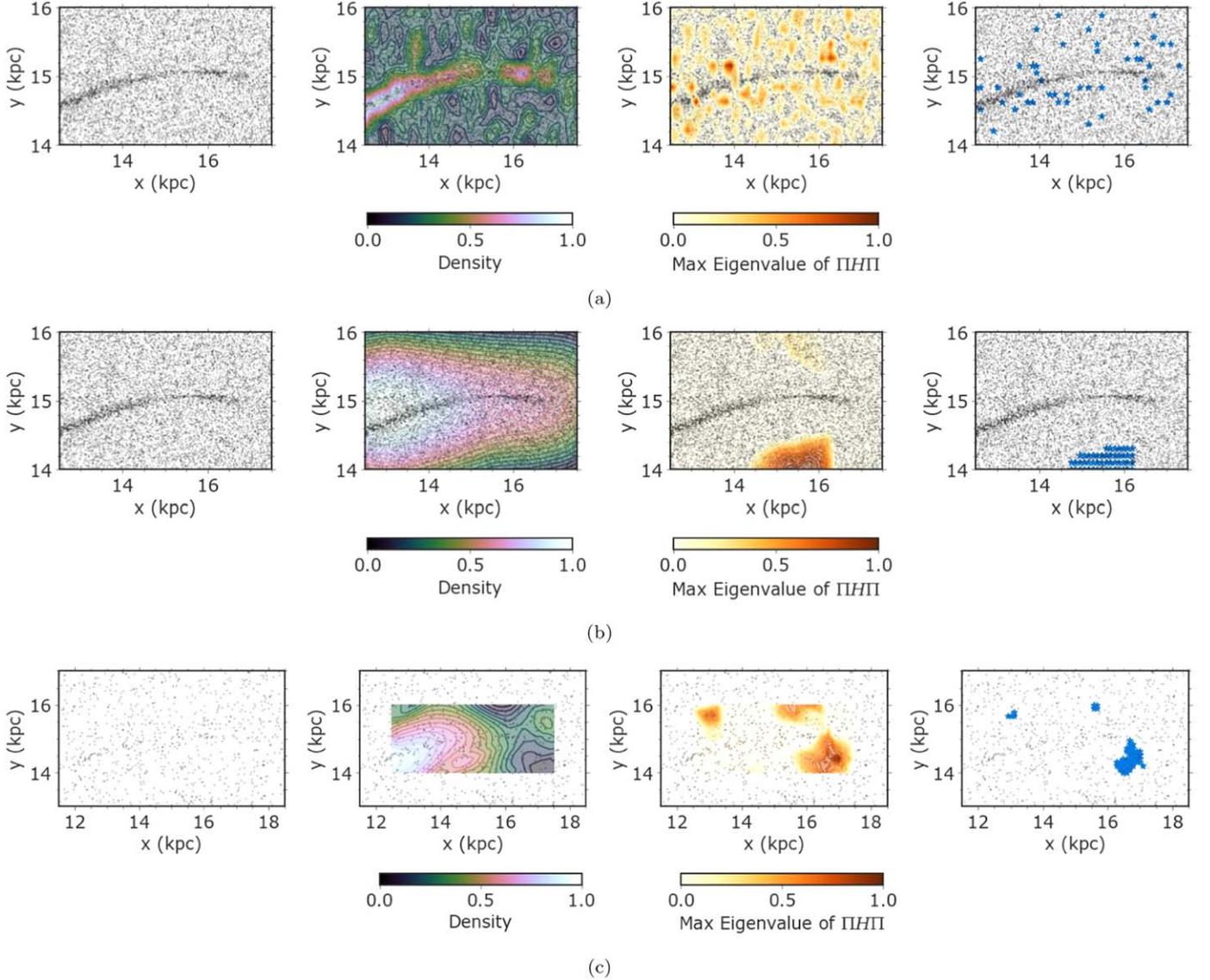

**Figure 12.** Illustration of the effect of the bandwidth on different modes of failures for our gap-detection pipeline. In the left panels, we show the simulations. In the center panels, filled-in contours show the map of the density and the maximum eigenvalues of $\Pi H \Pi$ that we used to locate gaps. We show gap indicator points in the right panels based on our percentile cuts. (a) Simulation of a stream at a distance of 0.8 Mpc using a bandwidth of 0.1 kpc for an exposure time of 1 hr. In this case, the predicted median location of gap points is close to the true location of the gap, but there are also spurious gaps in the background. Using our metrics defined in Section 4.4, this detection would have a small value for $\Delta$ but a large value for $\mathcal{S}_g$. (b) Simulation of a stream at a distance of 0.8 Mpc using a large bandwidth of 1.5 kpc for a 1 hr exposure time. The stream can still be identified, but the method could not detect the central gap. Using our pipeline selection metrics, this detection would be rejected on the basis that the spread in the locations of gap points is large and that the absolute deviation of the predicted gap location from the center is large, consistent with our $\Delta$ metric (see Section 4.4). (c) Simulation of a stream at 8.6 Mpc with a bandwidth of 0.8 kpc for a 1000 s exposure. Due to the very low density in the image, the gap-finder tool detects an off-centered gap but does not detect the real gap.

### F.3. Comparing Streams with Gaps to Intact Streams and Backgrounds

We compare the performance of the gap-detection tool to simulations of an intact stream with no perturbation from the dark matter subhalo for bandwidths of 0.5 and 0.8 kpc, and a stream with an off-centered gap with a bandwidth of 0.8 kpc in Figure 13. We followed the methodology highlighted in the main text to generate mock observations. Spurious gaps were persistent in the backgrounds, but careful visual inspection can rule out these detections. Additionally, the tool can detect off-centered gaps.

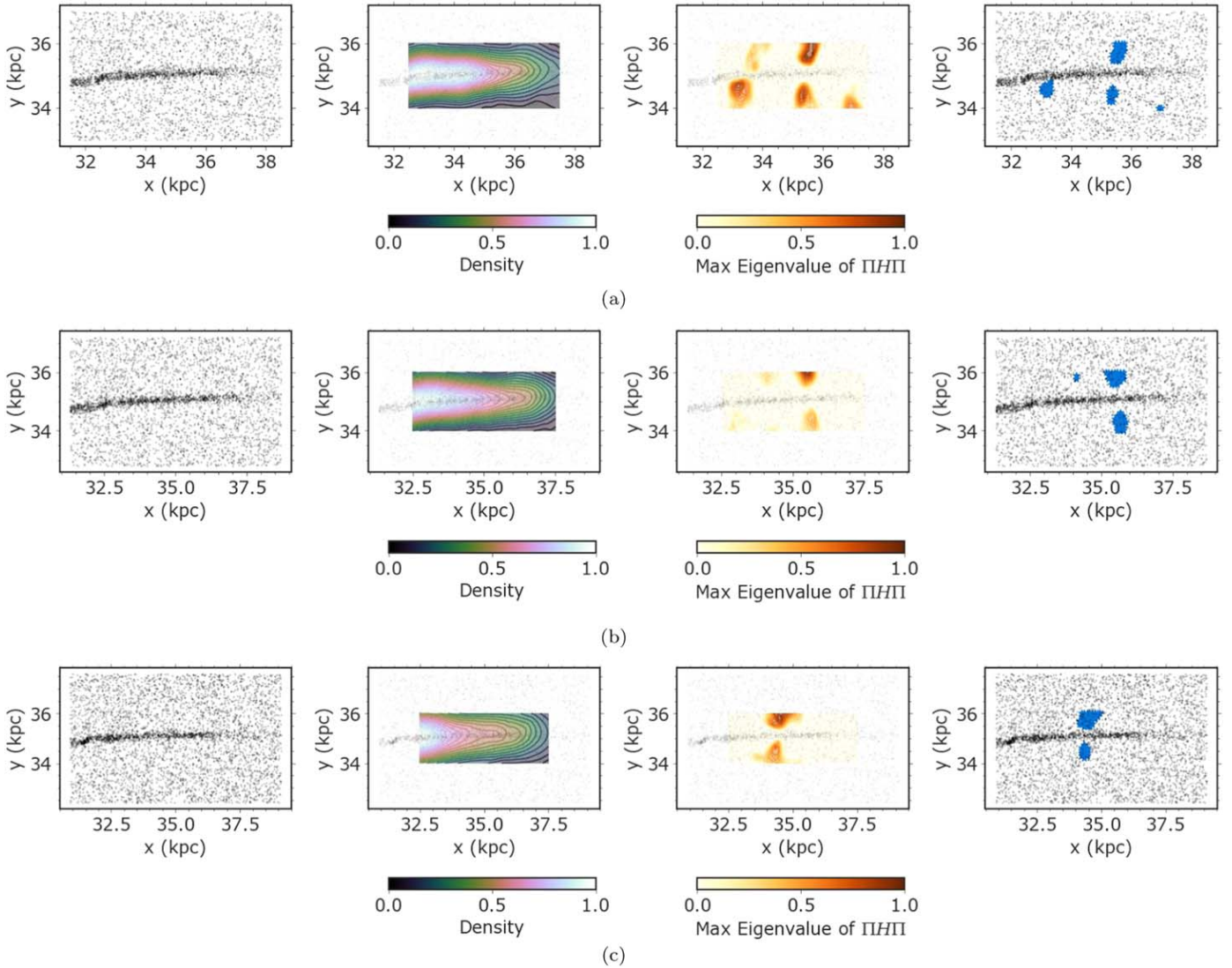

**Figure 13.** Additional tests of the gap-detection tool on intact streams for a bandwidth of 0.5 kpc (a) and a bandwidth of 0.8 kpc (b). We also show a stream with an off-centered gap (c) as an additional validation that our pipeline does not depend on the position of the gap. All simulated streams are at 1 Mpc at $R_{GC} = 35$ kpc for a 1000 s exposure.

## ORCID iDs

Christian Aganze  https://orcid.org/0000-0003-2094-9128
Sarah Pearson  https://orcid.org/0000-0003-0256-5446
Tjitske Starkenburg  https://orcid.org/0000-0003-2539-8206
Gabriella Contardo  https://orcid.org/0000-0002-3011-4784
Kathryn V. Johnston  https://orcid.org/0000-0001-6244-6727
Kiyan Tavangar  https://orcid.org/0000-0001-6584-6144
Adrian M. Price-Whelan  https://orcid.org/0000-0003-0872-7098
Adam J. Burgasser  https://orcid.org/0000-0002-6523-9536